\documentclass[journal]{IEEEtran}
\usepackage{multirow}
\usepackage{lineno}
\usepackage[nospace,noadjust]{cite}
\ifCLASSINFOpdf
\else
\usepackage[dvips]{graphicx}
\usepackage{color}
\fi
\usepackage[cmex10]{amsmath}
\usepackage{amssymb, amsmath}
\usepackage{amsmath}
\usepackage{graphicx}
\usepackage{multicol}
\usepackage{blindtext}
\usepackage{caption}
\usepackage{color, colortbl}
\definecolor{Gray}{gray}{0.9}

\begin{document}
\title{Multiframe-based Adaptive Despeckling Algorithm for Ultrasound B-mode Imaging with Superior Edge and Texture
\author{Jayanta~Dey, Md.~Kamrul~Hasan}
\thanks{Jayanta Dey and M. K. Hasan are with the Department
of Electrical and Electronic Engineering, Bangladesh University of
Engineering and Technology, Dhaka, Bangladesh (corresponding author's e-mail: jdey4@jhmi.edu).}}
\maketitle

\begin{abstract}
Removing speckle noise from medical ultrasound images while preserving image features without introducing artifact and distortion is a major challenge in ultrasound image restoration. In this paper, we propose a multiframe-based adaptive despeckling (MADS) algorithm to reconstruct a high-resolution B-mode image  from raw radio-frequency (RF) data that is based on a multiple input single output (MISO) model. As a prior step to despeckling, the speckle pattern in each frame is estimated using a novel multiframe-based adaptive approach for ultrasonic speckle noise estimation (MSNE) based on a single input multiple output (SIMO) modeling of consecutive deconvolved ultrasound image frames. The elegance of the proposed despeckling algorithm is that it addresses the despeckling problem by completely following the signal generation model unlike conventional ad-hoc smoothening or filtering based approaches, and therefore, it is likely to maximally preserve the image features. As deconvolution is a necessary pre-processing step to despeckling, we describe here a $2$-D extension of the SIMO model-based $1$-D deconvolution method. Finally, a complete framework for the generation of high-resolution ultrasound B-mode image has been also established in this paper. The results show $8.55-15.91$ dB, $8.24-14.94$ dB improvement in terms of SNR and PSNR, respectively, for simulation data and $2.22-3.17$, $13.24-32.85$ improvement in terms of NIQE and BRISQUE, respectively, for \textit{in-vivo} data compared to the traditional despeckling algorithms. Visual comparison shows superior texture, resolution, details of B-mode images offered by our method compared to those by a commercial scanner, and hence, it may significantly improve the diagnostic quality of ultrasound images. 

\begin{IEEEkeywords}
 Clinical ultrasound, analytic modeling, SIMO model, $2$-D deconvolution, MISO model, despeckling, high-resolution B-mode.
\end{IEEEkeywords}

\end{abstract}

\section{INTRODUCTION}
\IEEEPARstart{U}{ltrasound} (US) imaging system being non-invasive, non-ionizing, portable, and cost effective has become the most prevalent diagnostic tool among all the currently available imaging modalities, e.g., X-ray, magnetic resonance imaging, and computed tomography. However, imperfection of US imaging system design and the underlying physical phenomena related with US image acquisition give rise to low resolution and speckle noise that tend to reduce the image contrast, obscure and blur image details such as inclusion and small structure boundary, tissue texture and thereby, decrease the quality and reliability of medical ultrasound \cite{michailovich2006despeckling}. Speckle noise is a granular pattern inherent in any coherent imaging modalities \cite{choi2018speckle}, \cite{de2011real} similar to ultrasound imaging. It results from the constructive and destructive interferences of the reflected echos with different phases and amplitudes from the target at the receiving transducer. Removal of multiplicative speckle noise from US images is difficult due to the challenge of maintaining the precise texture of the image \cite{joel2018extensive}. As traditional despeckling filters distort the original image texture and introduce artifacts like blurring edges, changing the shape of structures present in the image by smoothening the noise corrupted image, the original noise affected images are sometimes more preferred than the noise-removed ones in the analysis where the image details have high importance \cite{joel2018extensive}. Therefore, an effective signal processing approach for removing speckle noise from the image while preserving the original tissue texture and small details of the image is vital to increase the diagnostic potential of medical ultrasound.

Speckle noise is generally modeled as multiplicative with the true noiseless image \cite{michailovich2006despeckling}, \cite{zong1998speckle}, \cite{achim2001novel}. However, it can be converted into the additive form by taking loagarithmic transformation of the noise-corrupted image. A number of algorithms have been reported in the literature for despeckling US images which attempts to remove speckle noise either in their multiplicative form or by converting the noise in the additive form. Among the first group of algorithms, the spatial averaging based approaches exploit the repetitive nature of the US image, and among them, linear filtering methods, such as Gaussian filter and mean filter are effective in reducing speckle noise \cite{choi2018speckle}. However, they tend to oversmooth the texture and blur edges present in the image. To overcome this problem, nonlinear approaches based on local \cite{frost1982model}, \cite{kuan1985adaptive} and non-local statistics \cite{coupe2009nonlocal}, \cite{buades2005review} of the image have been proposed. These algorithms are mainly weighted filters, in which the weights depend on the similarity between the intensity values of the patches surrounding the pixels \cite{shahdoosti2018maximum}. The main difference between local and non-local means methods is that the non-local means method employs the most similar pixels in the image to denoise the current pixel regardless of their Euclidean distance. Although these approaches tend to preserve textures and edges, their performance is dependent on tuning parameters, such as filter and patch size. Again, the nonlinear approaches based on the diffusion equation \cite{yu2002speckle} not only preserves edges but also enhances edges by inhibiting diffusion across edges and allowing diffusion on either side of the edge. However, selection of the  parameter-values is a major issue in this method, as a value of parameter that is smaller than the optimum one leads to unsatisfactory noise suppression whereas a higher parameter-value results in poor structure preservation \cite{jain2019non}.

The second group of algorithms deals with the speckle noise in its additive form. Now, as shown in \cite{michailovich2006despeckling}, the additive noise can be handled using any traditional denoising scheme, and the performance of this scheme determines the overall efficacy of despeckling. Finally, the denoised image is exponentially transformed back to give the despeckled image. The overall process is termed as homomorphic filtering approach \cite{michailovich2006despeckling}, and among the denoising schemes used in this process, the transfrom domain or multi-resolutional based appraoches \cite{gupta2005locally, gupta2005despeckling, yue2006nonlinear, koundal2016speckle} are of higher efficacy. In \cite{luisier2007new}, an advanced ultrasound despeckling algorithm is proposed based on the intra-scale correlation between the wavelet coefficients. Among the multi-resolutional approaches, as shown in \cite{joel2018extensive}, despeckling based on non-subsampled contourlet gives superior performance. However, in wavelet-based despeckling, a threshold is a critical parameter that is to be determined based on the \textit{a priori} knowledge of the distribution of the speckle pattern. In addition, the threshold-based filtering of wavelet coefficients implies texture smoothening, and it gives rise to artifact such as Gibbs phenomenon near the edges \cite{bhuiyan2007spatially}.  All of the approaches discussed so far relies on ad-hoc filtering or smoothening technique without addressing mathematically the speckle noise generation model. Therefore, it cannot be guaranteed that these algorithms only operate on the speckle noise without significantly distorting the true image.

In this paper, we propose a multiframe-based adaptive despeckling (MADS) algorithm that treats the speckle noise in its multiplicative form and utilizes the speckle patterns estimated using the multiframe-based adaptive ultrasonic speckle noise estimation (MSNE) algorithm proposed in this paper. The MSNE algorithm is based on formulating the true image as single input and the envelope of deconvolved consecutive US image frames with multiplicative speckle noise pattern in each frame as multiple outputs, i.e., a SIMO model \cite{dey2018ultrasonic}, \cite{hasan2016blind}. The despeckling algorithm, on the contrary, treats the envelope of deconvolved consecutive US image frames as multiple inputs and the true image as single output, i.e., a MISO model. According to the mathematical model representing the US imaging system, deconvolution is necessary prior to despeckling for resolution enhancement of the raw RF data. Hence, a $2$-D deconvolution approach as an extension of our previously published $1$-D deconvolution algorithm, i.e., \textit{b}MCFLMS \cite{dey2018ultrasonic} has been also described in this paper. To prevent misconvergence of the MSNE algorithm in the presence of additive noise and estimation error resulting from the deconvolution step, a zero-lag correlation contraint derived from the deconvolved image and the estimated speckle pattern is attached with the original cost function. As the overall despeckling approach has been derived by completely following the signal generation models, it is likely to maximally preserve the diagnostically important details and tissue texture present in the image. Finally, a complete framework including deconvolution, despeckling, and post-processing for ultrasound B-mode image generation with superior quality in terms of resolution, edges, small details and texture is also established in this paper for greater interest of the researchers.

The paper is organized as follows. Section II describes the US imaging model and formulates the problems to be addressed. The $2$-D deconvolution and speckle noise estimation based on SIMO models along with the derivation of a MISO model for despeckling the decovolved US images using the estimated speckle patterns are presented in Section III. Post-processing on the despeckled image is explained in Section IV. The performance of the proposed method is tested using simulation and \textit{in-vivo} data in Section V, and the pros and cons of the proposed framework is discussed in Section VI. Finally, summarizing the contributions with highlights for future research the paper concludes in Section VII.

\section{Problem Formulation}
Low resolution and speckle noise are the major issues related with US imaging. However, an overall US image enhancement  can be achieved by addressing these two issues in a sequential two-step process as described in \cite{michailovich2006despeckling}. First, the correlation between the image samples is to be minimized to increase the image resolution, and second, speckle noise has to be removed from the decorrelated image to improve the image contrast and better visualize the tissue texture. To achieve the aforementioned two objectives, a suitable model representing the US imaging system is necessary. With this in view, considering linear wave propagation through the tissue, and the scattering of the ultrasound pulse in the tissue as weak, we can use the first order Born approximation and consider the tissue scattering system as a linear system \cite{michailovich2006despeckling}. Therefore, the blurring, i.e., low resolution of an RF-image can be modeled as the result of convolution between the point-spread function (PSF) $s(m,n)$ of the imaging system with the tissue reflectivity function (TRF) $h(m,n)$ \cite{michailovich2006despeckling}, \cite{xue2018parametric}, \cite{al2018faster}. Mathematically, this can be written as
\begin{equation}\label{convolutionModel}
x(m,n) = s(m,n)*h(m,n) + v(m,n)
\end{equation}
where $x(m,n)$ is the backscattered ultrasound image data from the $n$-th A-line at discrete time $m$ and $v(m,n)$ is the additive noise associated with measurement error and other physical phenomena not accounted by the convolution model. In US imaging, $h(m,n)$ is corrupted by speckle noise, and as described in \cite{michailovich2006despeckling}, \cite{zong1998speckle}, \cite{achim2001novel}, $h_e(m,n)$, i.e., the envelope of $h(m,n)$ can be modeled as multiplicative with the true image as
\begin{equation}\label{speckleModel}
h_e(m,n) = r(m,n) u(m,n) + \zeta(m,n)
\end{equation}  
where $r(m,n)$, $u(m,n)$, and $\zeta(m,n)$ are true image, speckle noise, and additive noise resulting from the deconvolution process and the portion not accounted by the multiplicative process, respectively. Therefore, in ultrasound imaging, we are given the back-scattered data $x(m,n)$, and our objective is to design algorithms to estimate the TRF $h(m,n)$ from $x(m,n)$, and thereafter, obtain the despeckled image $r(m,n)$ from the envelope of $h(m,n)$.

\section{Method}
In this paper, our main concern is to derive a novel despeckling algorithm that is likely to preserve maximum features present in the image. However, according to \eqref{convolutionModel} and \eqref{speckleModel}, deconvolution is a necessary pre-processing step to despeckling. Therefore, for the completeness of a high-resolution B-mode image generation, we also consider it important to include a $2$-D extension of our previously proposed $1$-D deconvolution (\textit{b}MCFLMS) \cite{dey2018ultrasonic} algorithm and some post-processing techniques like gamma correction and grey level transformation so that the whole paradigm is described in a single paper.

\subsection{Deconvolution of RF Echo Data}
In this section, we attempt to estimate the TRF, $h(m,n)$, in \eqref{convolutionModel} with increased resolution by removing the effect of the PSF, $s(m,n)$, from the raw RF data, $x(m,n)$.
\begin{figure}[ht!]
\centering
\includegraphics[width = 3.3 in, height = 2.4 in]{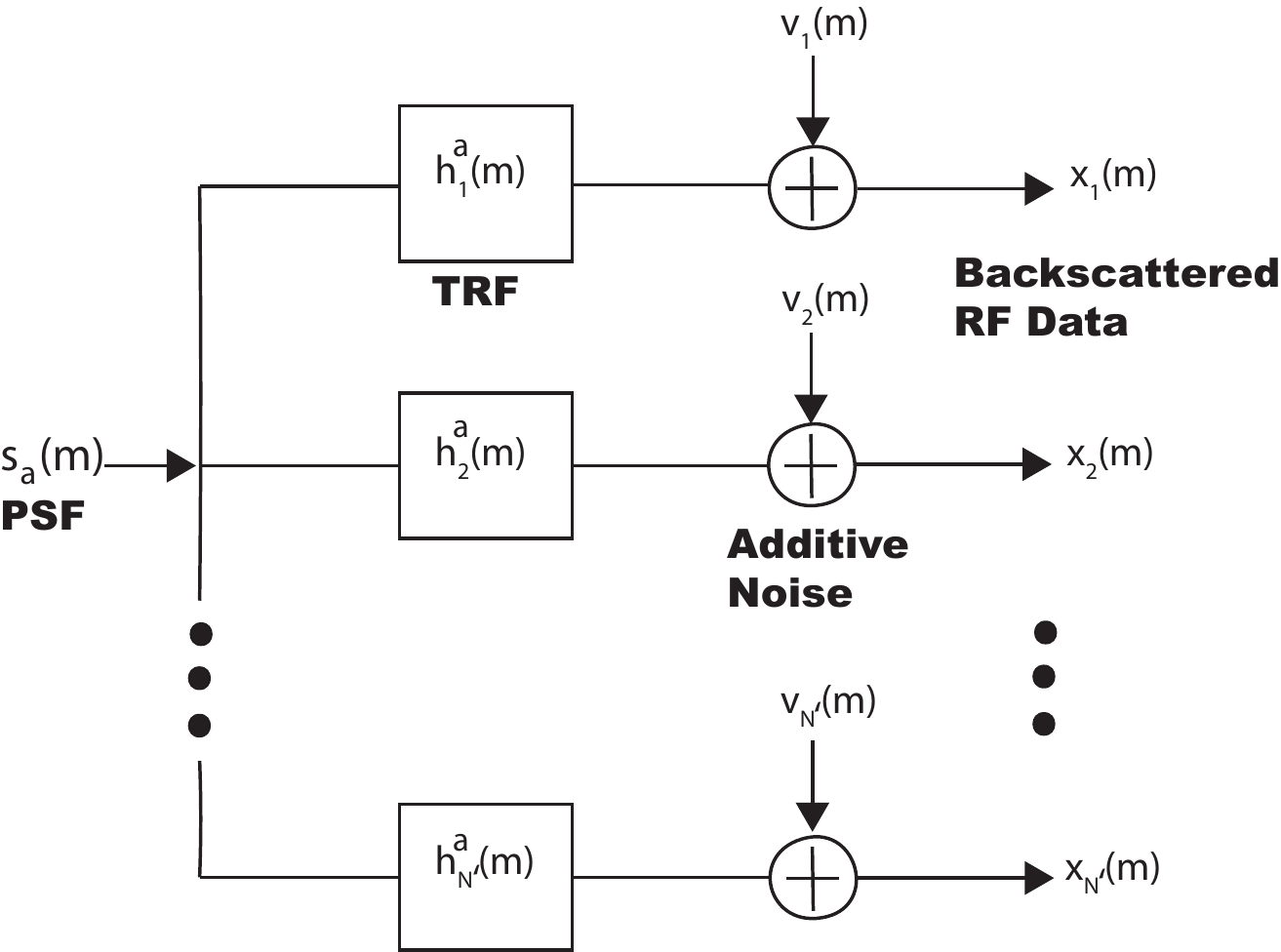}
 \caption{A SIMO model for backscattered RF signal.}
 \label{deconvolution}
\end{figure}
However, to estimate $h(m,n)$, similar to the approach as described in \cite{abeyratne1995higher}, we consider the $2$-D distortion kernal, i.e., PSF $s(m,n)$ in \eqref{convolutionModel} decomposible into two $1$-D distortion kernals (PSFs): one along the axial direction and the other along the lateral direction. Following this assumption, \eqref{convolutionModel} can be modified as
\begin{equation}
x(m,n) = s_a(m) *_a s_l(n) *_l h(m,n) + v(m,n)
\end{equation}
where $s_l(n)$ and $s_a(m)$ are the lateral and axial PSFs, respectively, and `$*_l$' and `$*_a$' represent convolution along the lateral and the axial directions, respectively. A novel technique for removing the effect of  axial PSF $s_a(m)$ from the measured RF image $x(m,n)$ was reported in our previous work \cite{dey2018ultrasonic} using a single input multiple output (SIMO) model as shown in Fig. \ref{deconvolution}, where $s_a(m)$  convolves with the $i$-th A-line of the axial TRF denoted as $h_i^a(m)$ and with additive noise $v_i(m)$ gives the $i$-th A-line RF data $x_i(m)$: 
\begin{equation}\label{axialConv}
x_i(m) = s_a(m) *_a h_i^a(m) + v_i(m)
\end{equation}
where
\begin{equation}
h_i^a(m) = s_l(n) *_l h(m,n) 
\end{equation}
In matrix form, \eqref{axialConv} can be written as
\begin{equation}
\mathbf{X} = \mathbf{S}_a \mathbf{h}^a + \mathbf{v}
\end{equation}
where $\mathbf{S}_a$ is the convolution matrix formed using the axial PSF $s_a(m)$ and
\begin{equation} \nonumber
\mathbf{X} = 
\begin{bmatrix}
\mathbf{X}_1 & \mathbf{X}_2 & \cdots & \mathbf{X}_{N'}
\end{bmatrix},
\end{equation}
\begin{equation} \nonumber
\mathbf{h}^a = 
\begin{bmatrix}
\mathbf{h}_1^a & \mathbf{h}_2^a & \cdots & \mathbf{h}_{N'}^a
\end{bmatrix}.
\end{equation}
Here, $N'$ is the total number of A-lines, and $\mathbf{X}_i$ and $\mathbf{h}_i^a$ are the $i$-th A-line with $M'$ samples taken from $x_i(m)$ and $h_i^a(m)$, respectively where $m = 1, 2, \cdots, M'$.
To account for the non-stationarity of the axial PSF, the RF data were divided into $B$ blocks with equal length $L_b$, and a block-based cost function $J^b$ for the $b$-th block was formulated (for details see \cite{dey2018ultrasonic}) to estimate the axial TRF block-by-block in the frequency-domain as
\begin{equation}
J^b = \sum_{i=1}^{n-1} \sum_{j=i+1}^n \tilde{\underline{\mathbf{e}}}_{ij}^{bH} \tilde{\underline{\mathbf{e}}}_{ij}^{b}
\end{equation}
where, `$H$' denotes the Hermitian operation, any variable with `$\underline{~~}$' represents the variable in the frequency-domain, and $\tilde{\underline{\mathbf{e}}}_{ij}^{b}$ is the Fourier transform of $\tilde{\mathbf{e}}_{ij}^{b}$ defined as
\begin{equation}\nonumber
\tilde{\mathbf{e}}_{ij}^b = \sum_{p=1}^{b-1} \mathbf{A}_1 \mathbf{e}_{ij}^{p(b-p)} + \sum_{p=1}^b \mathbf{A}_2 \mathbf{e}_{ij}^{p(b-p+1)}
\end{equation}
Here,
\begin{equation}\label{eqn:last_err}\nonumber
\mathbf{e}_{ij}^{p(b-p)} = \mathbf{C}_{\tilde{x}_i^p}\hat{\mathbf{h}}_j^{a(b-p)} - \mathbf{C}_{\tilde{x}_j^p}\hat{\mathbf{h}}_i^{a(b-p)} ,
p = 1,2,\cdots,b-1,
\end{equation}
\begin{equation}\label{eqn:first_err}\nonumber
\mathbf{e}_{ij}^{p(b-p+1)} = \mathbf{C}_{\tilde{x}_i^p}\hat{\mathbf{h}}_j^{a(b-p+1)} - \mathbf{C}_{\tilde{x}_j^p}\hat{\mathbf{h}}_i^{a(b-p+1)},
p = 1,2,\cdots,b,
\end{equation}
and $\mathbf{A}_1$, $\mathbf{A}_2$ are the truncation matrices truncating the last $(L_b-1)$ and the first $L_b$ samples of the error function, respectively. $\mathbf{C}_{\tilde{x}_i^p}$ is the convolution matrix formed using the RF data, $x(m,n)$ along the $i$-th A-line and the $p$-th block. Now, the $b$-th block axial TRF $\underline{\mathbf{h}}^{ab}$ was estimated as
\begin{equation}\label{eqn:arg}
\underline{\hat{\mathbf{h}}}^{ab}  = arg_{\underline{\mathbf{h}}^{ab}} \min J^b, \mbox{subject~to~}
||\underline{\hat{\mathbf{h}}}^a|| = 1
\end{equation}
where `$||\cdot||$' denotes the $l_2$-norm and
\begin{equation}\label{eqn:total_trf}
\underline{\hat{\mathbf{h}}}^a =
\begin{bmatrix}
\underline{\hat{\mathbf{h}}}^{a1T} & \underline{\hat{\mathbf{h}}}^{a2T} & \cdots  & \underline{\hat{\mathbf{h}}}^{aBT}
\end{bmatrix}^T
\end{equation}
with `$T$' denoting matrix transpose operation and
\begin{equation}
\underline{\hat{\mathbf{h}}}^{ab} =
\begin{bmatrix}
\underline{\hat{\mathbf{h}}}_{1}^{ab} & \underline{\hat{\mathbf{h}}}_{2}^{ab}  & \cdots  & \underline{\hat{\mathbf{h}}}_{N'}^{ab}
\end{bmatrix}
\end{equation}
Here, $\underline{\hat{\mathbf{h}}}_{i}^{ab}$ denotes the estimated $i$-th A-line $b$-th block axial TRF. In sample-domain, the estimated axial TRF along the $i$-th A-line can be written as $\hat{h}_i^a(m)$. Now, the estimated axial TRF $\hat{h}_i(m)$ along the $i$-th A-line at discrete time $m$ can be modeled as
\begin{equation}\label{lateralDeconv}
\hat{h}_i^a(m) = s_l(n) *_l h(m,n) + v'(m,n)
\end{equation}
where $v'(m,n)$ is the noise resulting from the estimation error of the axial TRF. Therefore, from \eqref{axialConv} and \eqref{lateralDeconv}, it is apparent that an attempt, similar to the approach adopted in the axial direction, can be made in the lateral direction to undo the effect of the lateral distortion kernal (PSF) from the estimated axial TRF $\hat{h}_i^a(m)$ to estimate the $2$-D deconvolved TRF $h(m,n)$.
In this approach, the PSF is considered to be laterally stationary as described in \cite{michailovich2005novel}, and therefore, no blocking is required in the lateral direction. The method is summarized in Table \ref{two-D-mcflms}.
\begin{table}[h!]  
  \caption{\textit{b}MCFLMS algorithm for $2$-D deconvolution of ultrasound RF image}
  \label{two-D-mcflms}
  \rule{8.5cm}{0.4pt}
  \begin{enumerate}
  \item[\bf{Step 1}] 
  \begin{enumerate}
  \item[.] Select the \textit{b}MCFLMS method reported in \cite{dey2018ultrasonic}
  \end{enumerate}
  \item[\bf{Step 2}]
  \begin{enumerate}
  \item[.] Set $\mathbf{X} =$ raw radio-frequency (RF) data, where $\mathbf{X}$ is the data to be deconvolved
  \item[.] Set the data length, $L = $ axial length, and the number of channels, $M = $ lateral length of the raw RF data. Execute the \textit{b}MCFLMS algorithm with block number, $B = 2$ along the axial direction.
   \end{enumerate}
   
   \item[\bf{Step 3}]
    \begin{enumerate}
    \item[.] Set $\mathbf{X} =$ axially deconvolved data in step 2
    \item[.] Set the data length, $L = $ lateral length, and the number of channels, $M = $ axial length of the raw radio-frequency (RF) data. Execute the \textit{b}MCFLMS algorithm with block number, $B = 1$ along the lateral direction.
    \end{enumerate}
 
  \end{enumerate}
  \rule{8.5cm}{0.4pt}
  \end{table}
The estimated TRF after lateral deconvolution is given by
\begin{equation}
\underline{\hat{\mathbf{h}}} =
\begin{bmatrix}
\underline{\hat{\mathbf{h}}}_1 & \underline{\hat{\mathbf{h}}}_2 & \cdots  & \underline{\hat{\mathbf{h}}}_{M'}
\end{bmatrix}^T
\end{equation}
where, $\underline{\hat{\mathbf{h}}}_i$ is the estimated lateral TRF along the $i$-th sample line, i.e., samples along the $i$-th row of $\underline{\hat{\mathbf{h}}}$. In sample-domain, the $m$-th row and $n$-th column sample of the estimated TRF, $\underline{\hat{\mathbf{h}}}$ can be denoted as $\hat{h}(m,n)$.
\subsection{Proposed Despeckling Algorithm} \label{sssec:ref1}
The envelope of the estimated TRF $\hat{h}(m,n)$ in the previous subsection is corrupted with speckle noise as given by \eqref{speckleModel}. It is apparent from \eqref{speckleModel} that knowledge of the speckle pattern $u(m,n)$ in an image frame can help despeckling that frame. However, in the absence of additive noise, direct division of $h_e(m,n)$ by $u(m,n)$ may amplify noise and/or give rise to division by zero problem. Hence, in this subsection, we attempt to formulate an energy constrained iterative approach to find an speckle noise cancelling multiplying factor to despeckle the frame. In what follows, we attempt to formulate a novel SIMO model for the deconvolved image frames and thereby, estimate the speckle noise in the respective deconvolved frames using an adaptive filtering technique. Then a novel MISO model will be proposed to despeckle the decconvolved image frames using the estimated speckle pattern of the respective frames.

\begin{figure}[ht!]
\centering
\includegraphics[width = 3.5 in, height = 2.2 in]{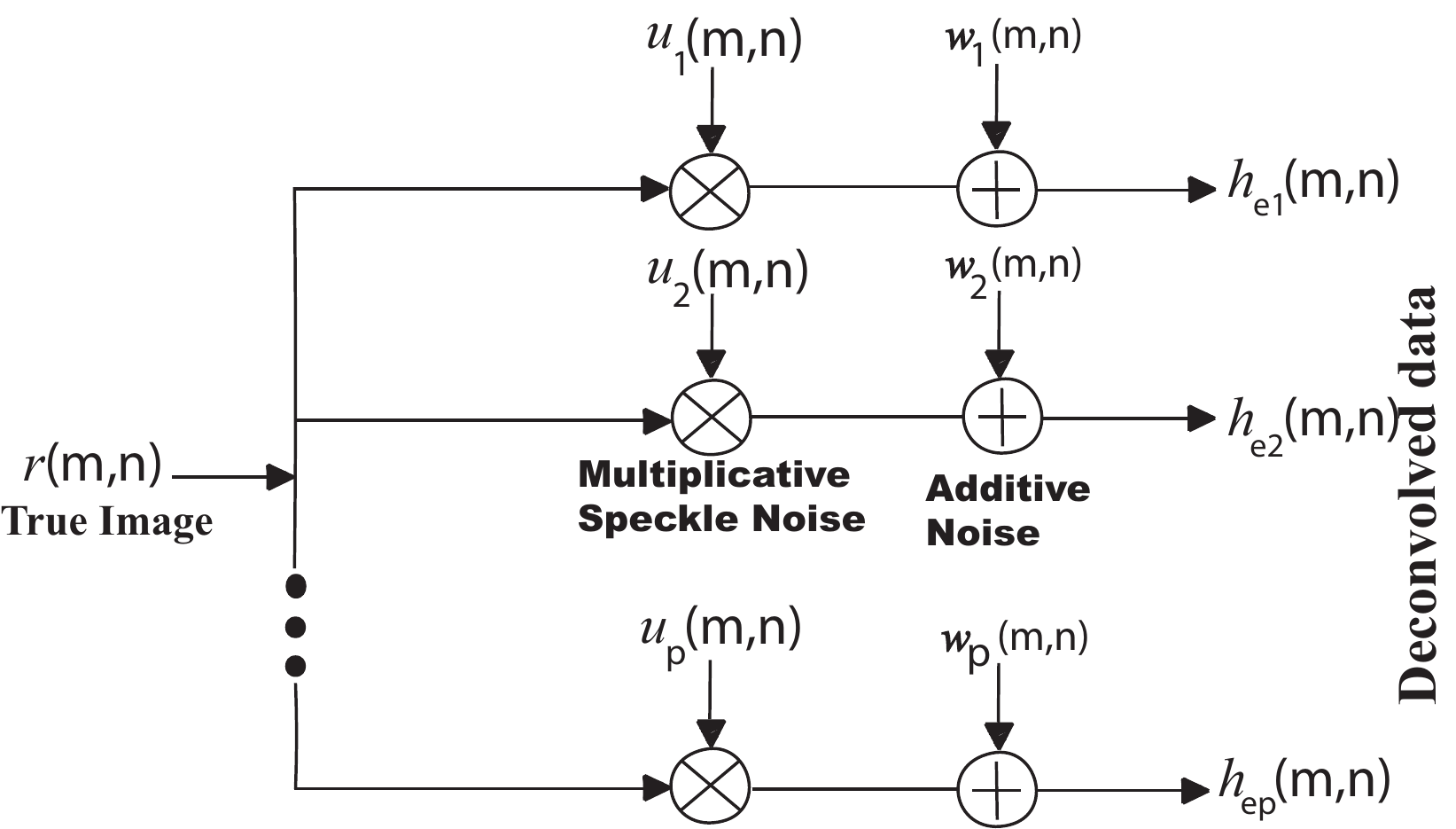}
 \caption{A new SIMO model for deconvolved $2$-D RF data.}
 \label{speckle}
\end{figure}
In an ultrasound imaging system, images are generally acquired at a frame rate ranging from $10-60$ frames per second (FPS) with speckle patterns generated randomly in each image frame from the interference of the US pulse at the receiving transducer. In our work, we attempt to use $p$ consecutive deconvolved frames to formulate a single input multiple output (SIMO) model as shown in Fig. \ref{speckle}. Here, the true ultrasound image $r(m,n)$ that is considered stationary throughout the $p$ frames, multiplies with the speckle noise $u_i(m,n)$ of the $i$-th frame, and with an additive noise $w_i(m,n)$ gives the envelope, $\hat{h}_{ei}(m,n)$, of the estimated deconvolved image $\hat{h}_i(m,n)$ of the $i$-th frame: 
\begin{equation} \label{simoBasic}
\hat{h}_{ei}(m,n) = r(m,n) u_i(m,n) + w_i(m,n)
\end{equation}
In matrix form, \eqref{simoBasic} can written as

\begin{equation}
\hat{\mathbf{H}}_{ei} = \mathbf{R}\cdot*\mathbf{U}_i + \mathbf{W}_i
\end{equation}
where `$\cdot*$' denotes elementwise multiplication. Here, $\hat{h}_{ei}(m,n)$, $r(m,n)$, $u_i(m,n)$ and $w_i(m,n)$ represents the $m$-th row and $n$-th column elements of $\hat{\mathbf{H}}_{ei}$, $\mathbf{R}$, $\mathbf{U}_i$ and $\mathbf{W}_i$, respectively. Therefore, here the challenge is to estimate the speckle noise $u_i(m,n)$ from each frame in the presence of additive noise $w_i(m,n)$ and then remove the speckle noise from the deconvolved envelope image $\hat{h}_{ei}(m,n)$.
The assumptions behind the SIMO model formulation and the identifiability condition \cite{xu1995least} for the speckle noise pattern $u_i(m,n)$ of the $i$-th frame are
\begin{enumerate}
\item[1.] The true image $r(m,n)$ is stationary throughout the $p$ consecutive frames.
\item[2.] The speckle patterns of each frame do not share common zeroes with the rest $p-1$ consecutive frames.
\end{enumerate}
These assumptions are realistic because for an ultrasound video recording with $30$ FPS, consecutive $5-10$ frames take around $0.17-0.33~$second during which the hand motion can be ignored. Then we can consider the true ultrasound image $r(m,n)$ as stationary throughout these frames. Again, formation of speckle noise in consecutive frames is a completely random phenomena, and hence, they are unlikely to contain common zeros. However, the probablity of sharing common zeros between the speckle patterns of the consecutive frames can be further reduced by increasing the number of fames into consideration; doing this will improve the identification accuracy of the patterns provided that the stationarity assumption remains valid as shown in the result section later. In what follows, we derive a multiframe-based adaptive speckle noise estimation algorithm using the proposed SIMO model.

\subsubsection{Speckle Noise Estimation}
In the absence of additive noise, the following error function $e_{ij}(m,n)$ can be used to estimate the speckle noise:
\begin{equation}\label{error}
e_{ij}(m,n) = \hat{h}_{ei}(m,n) \hat{u}_j(m,n) - \hat{h}_{ej}(m,n) \hat{u}_i(m,n)
\end{equation}
where $\hat{u}_i(m,n)$ is the estimated speckle noise of the $i$-th frame. Notice that for additive noiseless case if we can estimate the speckle pattern accurately, the error function defined in \eqref{error} becomes zero. Using this fact, we can build the following cost function to iteratively estimate the speckle noise:

\begin{equation}\label{costFunction}
J = \sum_{i=1}^{p-1} \sum_{j = i+1}^p ||\mathbf{E}_{ij}\cdot* \mathbf{E}_{ij}||_F^2
\end{equation}
where
\begin{equation}
\mathbf{E}_{ij} = \hat{\mathbf{H}}_{ei}\cdot* \hat{\mathbf{U}}_j - \hat{\mathbf{H}}_{ej}\cdot*\hat{\mathbf{U}}_i
\end{equation}
and `$||\cdot||_F$' indicates the Frobenius norm. An estimate of the speckle noise $\hat{\mathbf{U}}$ can be obtained by minimizing the cost function $J$ as

\begin{equation} \label{estimateU}
\hat{\mathbf{U}}  = arg_{\mathbf{U}} \min J, \mbox{subject~to~}||\hat{\mathbf{U}}||_F = 1
\end{equation}
where
\begin{equation}\label{s}
\hat{\mathbf{U}} =
\begin{bmatrix}
\hat{\mathbf{U}}_1 & \hat{\mathbf{U}}_2  & \cdots  & \hat{\mathbf{U}}_p
\end{bmatrix}
\end{equation}
Taking the gradient of $J$ in \eqref{costFunction}, we get
\begin{equation} \label{gradient}
\nabla_k J = \frac{\partial J}{\partial \hat{\mathbf{U}}_k} = 2\sum_{i=1}^p \hat{\mathbf{H}}_{ei}.* \mathbf{E}_{ik}
\end{equation}
The update equation for the MSNE algorithm at the $q$-th iteration is

\begin{equation}\label{update}
\hat{\mathbf{U}}(q+1) = \hat{\mathbf{U}}(q) - \mu(q) \nabla J(q)\big|_{\mathbf{U}=\hat{\mathbf{U}}(q)}
\end{equation}
where,
\begin{align}
\nabla J(q) &= \frac{\partial J(q)}{\partial \hat{\mathbf{U}}} \nonumber\\
&=
\begin{bmatrix}
\nabla_1 J(q) & \nabla_2 J(q)  & \cdots  & \nabla_p J(q)
\end{bmatrix}
\end{align}
and $\mu(q)$ is the variable step-size (VSS) which is such that the misalignment of $\hat{\mathbf{U}}(q+1)$ with the true noise pattern $\mathbf{U}$ is minimum at every iteration, given the current estimate $\hat{\mathbf{U}}(q)$:
\begin{align}\label{muUp}
J_\mu (q) &= (|| \mathbf{U} - \alpha \hat{\mathbf{U}}(q+1) ||_F)^2| \hat{\mathbf{U}}(q) \nonumber\\
		  &= \big|\big| \mathbf{U}\cdot*\mathbf{U} - 2\alpha \mathbf{U}\cdot*\hat{\mathbf{U}}(q) + 2\alpha\mu (q)\mathbf{U}\cdot*\nabla J(q) \nonumber \\
		  & + \alpha^2\hat{\mathbf{U}}(q)\cdot*\hat{\mathbf{U}}(q) - 2\alpha^2\mu (q)\hat{\mathbf{U}}\cdot*\nabla J(q) \nonumber\\
		  & + \alpha^2\mu^2 (q) \nabla J(q)\cdot*\nabla J(q) \big|\big|_S | \hat{\mathbf{U}}(q)
\end{align}
where $\alpha$ is a scaling constant inherent in any blind channel identification approach based on the cross-relation, and we define an operator `$||\cdot||_S$' which evaluates the sum of the matrix elements. Minimizing \eqref{muUp}, i.e., setting the gradient of $J_\mu(q)$ with respect to $\mu(q)$ to zero, we get $\mu(q)$ as 
\begin{equation}\label{vss}
\mu(q) = \frac{\big|\big|\hat{\mathbf{U}}(q)\cdot*\nabla J(q)\big|\big|_S}{\big|\big|\nabla J(q)\cdot*\nabla J(q)\big|\big|_S}
\end{equation}
Equation \eqref{vss} can be considered as a variant of VSS derived in \cite{gaubitch2006generalized}.

Thus far additive noise has been ignored in the derivation of the proposed MSNE algorithm. However, it has a similar effect on the convergence of the MSNE algorithm as described in \cite{dey2018ultrasonic}. To solve the problem, we need to impose a constraint on \eqref{costFunction} so as to prevent the deviation of the estimated speckle pattern from the true speckle pattern. To this end, consider the following model of speckle corrupted image for the $k$-th frame as described in \cite{coupe2009nonlocal}:
\begin{equation}\label{simEqn}
h_{ek}(m,n) = r(m,n) + r^\eta(m,n)~\mathcal{V}_k(m,n)
\end{equation}
where $\mathcal{V}_k(m,n) \sim \mathcal{N}(0,~\sigma^2)$. Here, \eqref{simEqn} models the random constructive and destructive interference phenomenon by adding the true image $r(m,n)$ with the non-linearly reflected image $r^\eta(m,n)$ having its phase and amplitude randomly altered by white noise $\nu_k(m,n)$. However, as discussed in \cite{coupe2009nonlocal}, $\eta = 0.5$ fits the data better. Ignoring the additive noise in \eqref{speckleModel} and comparing with \eqref{simEqn}, our proposed algorithm is basically estimating the true speckle pattern as
\begin{equation}\label{trueSpeckle}
u_k(m,n) = 1 + r^{(\eta-1)}(m,n)~\mathcal{V}_k(m,n)
\end{equation}
Then the speckle pattern $\hat{u}_k(m,n)$, estimated using the proposed method, can be expresses as
\begin{equation}\label{estimatedSpeckle}
\hat{u}_k(m,n) = 1 + r^{(\eta-1)}(m,n)~\hat{\mathcal{V}}_k(m,n)
\end{equation}
Since $r(m,n)$ is not a variable here, to make the estimated speckle pattern $\hat{u}_k(m,n)$ in \eqref{estimatedSpeckle} close to the true speckle pattern $u_k(m,n)$ in \eqref{trueSpeckle}, we need to maximize the zero-lag correlation between $\hat{\mathcal{V}}_k(m,n)$ and $\mathcal{V}_k(m,n)$. However, to attach this as a constraint on \eqref{costFunction}, estimates of the true image $r(m,n)$ and the parameter $\eta$ are necessary. Alternatively, consider the zero-lag correlation between the deconvolved image $\hat{h}_{ek}(m,n)$ and the estimated speckle pattern $\hat{u}_k(m,n)$:
\begin{align}\label{constraint}
J_{corr} &= \big| \big| \hat{\mathbf{H}}_{ek}\cdot*\hat{\mathbf{U}}_k \big| \big|_S \nonumber \\
         &= \big| \big| (\mathbf{H}_{ek} + \mathbf{W}_k)\cdot*\hat{\mathbf{U}}_k \big| \big|_S \nonumber \\
         &= \big| \big| \mathbf{H}_{ek}\cdot*\hat{\mathbf{U}}_k \big| \big|_S
\end{align}
where the zero-lag correlation between the additive noise $\mathbf{W}_k$ and speckle noise $\mathbf{U}_k$ is considered zero. Using \eqref{simEqn} and \eqref{estimatedSpeckle} in \eqref{constraint}, we get
\begin{align}\label{show}
J_{corr} &= \big| \big| \mathbf{R} + \mathbf{R}^\eta\cdot*\mathcal{V}_k + \mathbf{R}^\eta\cdot*\hat{\mathcal{V}}_k + \mathcal{V}_k\cdot* \hat{\mathcal{V}}_k \big| \big|_S \nonumber\\
         &= \big| \big| \mathbf{H}_{ek} + \mathbf{R}^\eta\cdot*\hat{\mathcal{V}}_k + \mathcal{V}_k\cdot* \hat{\mathcal{V}}_k \big| \big|_S \nonumber\\
         &= c + \big| \big| \mathcal{V}_k\cdot* \hat{\mathcal{V}}_k \big| \big|_S
\end{align}
where at the point of misconvergence when the estimated speckle pattern is close to the true speckle pattern, the zero-lag correlation between $r^\eta(m,n)$ and $\hat{\mathcal{V}}_k(m,n)$ can be considered zero, and $c$ is a constant defined as $c = \big| \big| \mathbf{H}_{ek} \big| \big|_S$.
From \eqref{show}, it is apparent that maximizing $J_{corr}$ or equivalently minimizing $-J_{corr}$ is analogous to maximizing the zero-lag correlation between $\hat{\mathcal{V}}_k(m,n)$ and $\mathcal{V}_k(m,n)$.
To prevent the misconvergence of the proposed algorithm in the noisy case, we propose to use the zero-lag correlation constraint in \eqref{show} with the MSNE cost function in \eqref{costFunction}. Then modifying \eqref{costFunction} for noisy case, we obtain
\begin{equation}\label{constraint2}
J_c(q) = J(q) - \beta_1 J_{corr}(q)
\end{equation}
where $\beta_1$ is the Lagrange multiplier or also known as the coupling factor. Taking gradient of \eqref{constraint2} with respect to $\hat{\mathbf{U}}_k$, we get
\begin{equation}\label{gradientConstraint}
\nabla_k J_c(q) = \nabla_k J(q) - \beta_1 \mathbf{H}_{ek}
\end{equation}
Replacing $\nabla_k J(q)$ in \eqref{gradient} by $\nabla_k J_c(q)$, we can estimate the speckle noise pattern in each of the $p$ frames.

\subsubsection{Estimation of the True Ultrasound Image}
\begin{figure}[ht!]
\centering
\includegraphics[width = 3.5 in, height = 2.6 in]{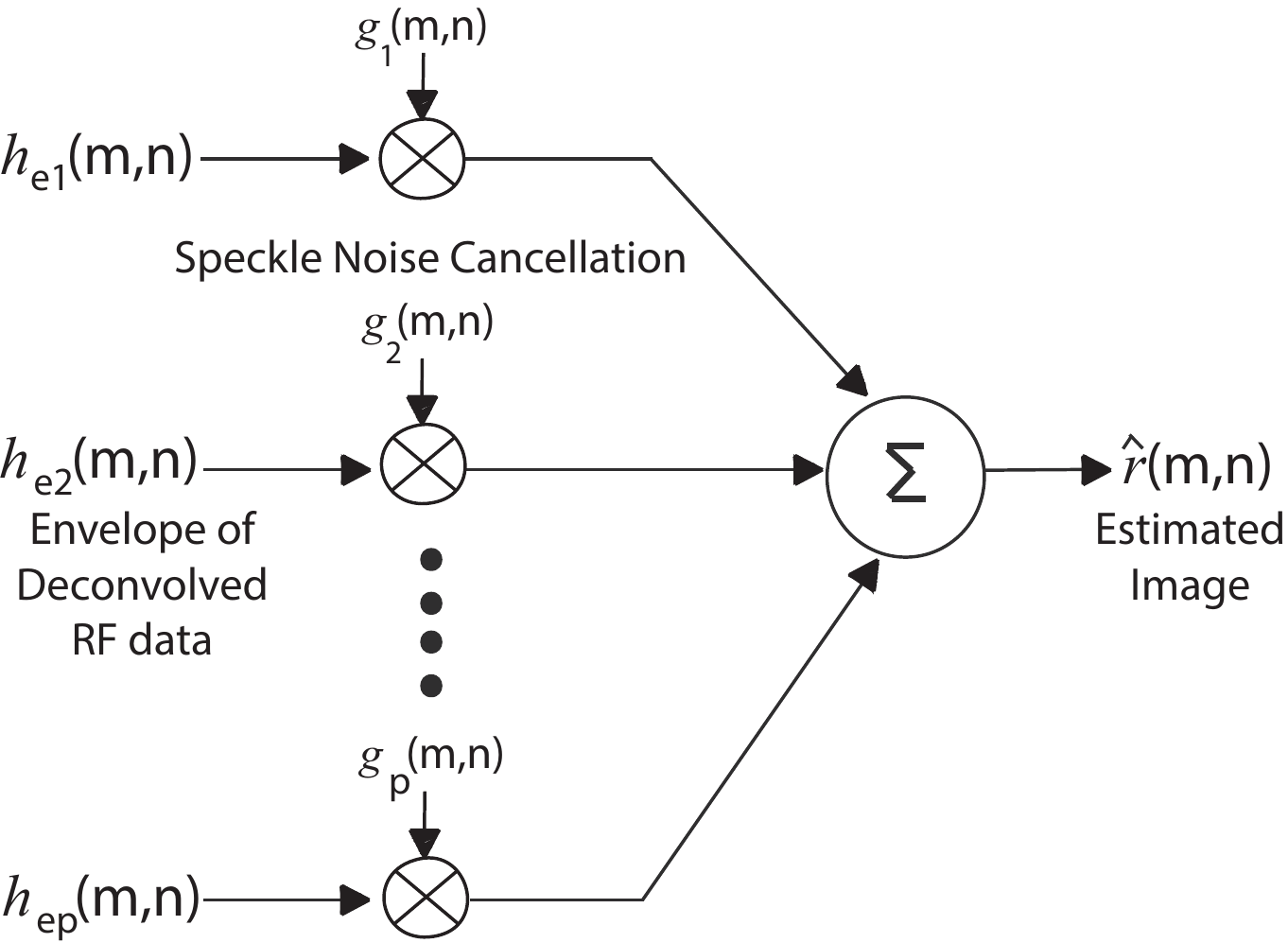}
 \caption{The block diagram for the estimation procedure of true ultrasound image.}
 \label{m-MINT}
\end{figure}

Hitherto, a novel algorithm for estimating the speckle pattern has been explained with a view to estimating the true ultrasound image $r(m,n)$ in \eqref{speckleModel} using $\hat{\mathbf{U}}$ from \eqref{estimateU}. Now, we describe a novel MISO model as shown in Fig. \ref{m-MINT}, where the speckle noise cancellation (SNC) factors for each of the $i$-th frame $g_i(m,n)$ multiplies with $h_{ei}(m,n)$, to obtain an estimate of the true US image $\hat{r}_i(m,n)$ for that frame. In the absense of additive noise in \eqref{simoBasic} and estimation error in the estimated speckle pattern $\hat{\mathbf{U}}_i$, the estimated ultrasound image of the $i$-th frame, $\hat{\mathbf{R}}_i$, can be obtained in matrix form as
\begin{align}
\hat{\mathbf{R}}_i &= \mathbf{H}_{ei}\cdot*\mathbf{G}_i \label{equalize}\\
				    &= \mathbf{R}_i\cdot*\mathbf{U}_i\cdot*\mathbf{G}_i \label{simpleEqualize}
\end{align}
where $\mathbf{G}_i$ is the elementwise multiplying SNC factor to equalize the speckle pattern for the $i$-th frame. Here for an accurate estimation of $\mathbf{R}_i$, the elements of the matrix $\mathbf{U}_i.*\mathbf{G}_i$ in \eqref{simpleEqualize} should be equal to $1$. Therefore, we can formulate the following cost function to estimate the SNC factors:
\begin{equation}\label{MINTcost}
J_{eq} = ||\mathbf{U}\cdot*\mathbf{G} - \mathbf{D}||_F^2
\end{equation}
where
\begin{equation}\label{g}
\mathbf{G} =
\begin{bmatrix}
\mathbf{G}_1 & \mathbf{G}_2  & \cdots  & \mathbf{G}_p
\end{bmatrix}
\end{equation}
and $\mathbf{D}$ is a matrix with all entries equal to $1$. In what follows, we describe the optimization of \eqref{MINTcost} to estimate the SNC matix $\mathbf{G}$. For simplicity, we first derive the adaptive algorithm considering the true speckle pattern $\mathbf{U}$ to avoid any estimation error. Then we adopt energy regularization and gradient averaging methods to account for the estimation error in $\hat{\mathbf{U}}$.
Taking gradient of \eqref{MINTcost} with respect to $\mathbf{G}$, we get
\begin{equation} \label{gradMINT}
\nabla J_{eq} = \frac{\partial J_{eq}}{\partial \mathbf{G}} = 2(\mathbf{U}.*\mathbf{G} - \mathbf{D}).*\mathbf{U}
\end{equation}
The update equation for estimating $\mathbf{G}$ at the $q'$-th iteration is given by
\begin{equation}\label{updateMINT}
\mathbf{G}(q'+1) = \mathbf{G}(q') - \mu(q') \nabla J_{eq}(q')
\end{equation}
where $\mu(q')$ is the VSS for the $q'$-th iteration, and can be obtained following \eqref{vss} as
\begin{equation}\label{mu}
\mu(q') = \frac{\big|\big|\mathbf{G}(q')\cdot*\nabla J_{eq}(q')\big|\big|_{S}}{\big|\big|~\nabla J_{eq}(q')\cdot*\nabla J_{eq}(q')\big|\big|_S}
\end{equation}
Up to now, we have ignored the effect of estimation error in $\hat{\mathbf{U}}$. If we assume that $\mathbf{E}_s$ be the estimation error in $\hat{\mathbf{U}}$, then we can write
\begin{equation}\label{errorU}
\hat{\mathbf{U}} = \mathbf{U} + \mathbf{E}_s
\end{equation}
Now, replacing the true speckle pattern $\mathbf{U}$ with the estimated speckle pattern $\hat{\mathbf{U}}$, the cost function in \eqref{MINTcost} becomes
\begin{align}
J_{eq} &= ||\hat{\mathbf{U}}.*\mathbf{G} - \mathbf{D}||_F^2 \label{hatEqualize}\\
       &= ||(\mathbf{U}+\mathbf{E}_s).*\mathbf{G} - \mathbf{D}||_F^2 \nonumber\\
	   &= ||\mathbf{U}.*\mathbf{G} - \mathbf{D}||_F^2 + \mathcal{E}
\end{align}
where $\mathcal{E}$ represents the terms including the estimation error $\mathbf{E}_s$. Therefore, the gradient in \eqref{gradMINT} will have two components-- one from the desired part of the cost function and the other from the estimation error part, i.e.,
\begin{equation} \label{gradMINT2}
\nabla J_{eq}(q') = \nabla J_{eq}^{desired}(q') + \nabla J_{eq}^{error}(q')
\end{equation}
From \eqref{hatEqualize}, we observe that $\hat{\mathbf{U}}$ and $\mathbf{G}$ have inverse relation so as to make their elementwise product equal to $1$. Therefore, in \eqref{errorU}, the term associated with the estimation error in $\hat{\mathbf{U}}$ gives rise to an SNC factor $\mathbf{G}$ in which the small estimation error is magnified. To make $\mathbf{G}$ less sensitive to such phenomenon, we impose an energy regularization constraint on $\mathbf{G}$ in \eqref{MINTcost}:
\begin{equation}\label{modifiedCost}
J'_{eq} = ||\hat{\mathbf{U}}\cdot*\mathbf{G} - \mathbf{D}||_F^2 + \beta_2||\mathbf{G}||_F^2
\end{equation}
where $\beta_2$ is the Lagrange multiplier. Now, the gradient in \eqref{gradMINT} becomes
\begin{equation} \label{gradMINT3}
\nabla J'_{eq} = 2(\hat{\mathbf{U}}\cdot*\mathbf{G} - \mathbf{D})\cdot*\hat{\mathbf{U}} + 2 \beta_2\mathbf{G}
\end{equation}
In addition to the energy regularization constraint, we propose the following gradient averaging technique to average out or at least reduce the detrimental effect of $\nabla J_{eq}^{error}(q')$:
\begin{equation}
\nabla J''_{eq}(q') = \alpha \nabla J'_{eq}(q') + (1 - \alpha) J'_{eq}(q'-1)
\end{equation}
where $\alpha$ is the weighting factor given on the current gradient $\nabla J_{eq}(q')$. Now, using the average gradient $\nabla J''_{eq}(q')$ in \eqref{updateMINT} and \eqref{mu} in place of $\nabla J_{eq}(q')$, we can estimate the SNC factor $\mathbf{G}$ which can then be used to get an estimate of the estimated true image of the $i$-th frame, $\hat{\mathbf{R}}_i$, using \eqref{equalize}. Finally, averaging the estimates for $i=1,2,\cdots,p$ for SNR improvement, the true image is reconstructed as
\begin{equation}\label{speckleless}
\hat{\mathbf{R}} = \frac{1}{p}\sum_{i=1}^p \hat{\mathbf{R}}_i
\end{equation}

The additive noise in \eqref{simoBasic} has been ignored so far. However, considering the additive noise in \eqref{equalize}, we get
\begin{align}\label{noiseAdder}
\hat{\mathbf{R}}_i &= (\mathbf{R}\cdot*\mathbf{U}_i + \mathbf{W}_i)\cdot*\mathbf{G}_i \nonumber\\
                    &= \hat{\mathbf{R}}'_i + \mathbf{W}'_i
\end{align}
where $\mathbf{W}'_i$ is the modified additive noise in the $i$-th frame.
From \eqref{speckleless} and \eqref{noiseAdder}, we can write
\begin{equation}\label{finalImg}
    \hat{\mathbf{R}} = \hat{\mathbf{R}}' + \mathbf{W}'
\end{equation}
Due to superior performance of  non-subsampled shearlet transform (NSST) to capture the geometric and mathematical properties of an image such as scales, directionality, elongated shapes and oscillations as described in \cite{shahdoosti2016image}, we attempt to denoise the image $\hat{\mathbf{R}}$ with a hard-thresholding on the NSST co-efficients following a similar method as described in \cite{gupta2014speckle}. Here, $\mathbf{W}'$ is somewhat minimized due to averaging. However, according to \cite{michailovich2006despeckling}, \cite{achim2001novel}, \cite{garg2018despeckling}, the effect of speckle noise is more pronounced compared to the additive noise and hence, the SNR for the additive noise can be considered high in the despeckled image. In this approach, the estimated true image is decomposed into $4$ levels with each having $3$, $3$, $4$ and $4$ directions. The coarse scales of the NSST co-efficients are not thresholded, but the finest scale is hard-thresholded using a tunable low threshold value.

\section{Post-processing}
\begin{figure}[ht!]
\centering
\includegraphics[width = 3.5 in, height = 2.6 in]{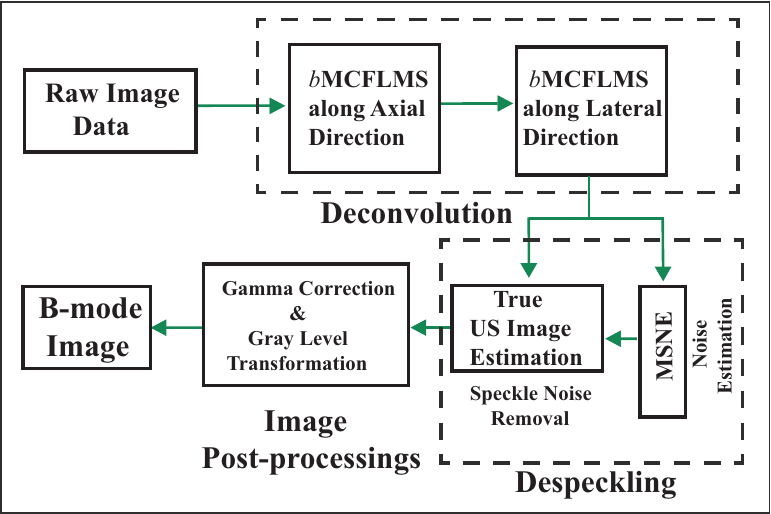}
 \caption{Block diagram of the proposed ultrasound image reconstruction method from the raw RF data.}
 \label{Block}
\end{figure}

To match with the characteristics of the display monitor and control the overall brightness of the image, further post-processing like gamma correction \cite{lee2005adjustable} is necessary. The gamma correction \cite{huang2018speckle} of the estimated true image $\hat{\mathbf{R}}$ is done using
\begin{equation}\label{gamma}
    I(m,n) = \left(\frac{\hat{r}(m,n)}{max(\hat{\mathbf{R}})}\right)^{\gamma}
\end{equation}
Finally, to control the image contrast, gray level transformation \cite{suetens2002fundamentals} of the image $I(m,n)$ is done as
\begin{equation}\label{gray}
    G(m,n) =
    \begin{cases}
      0, \hfill \quad \quad \quad \quad \quad \quad \quad \quad \quad \quad  \text{if}\ I(m,n) < W_{low} \\
      \frac{(I(m,n)/max(\mathbf{I})-W_{low})}{(W_{high} - W_{low})}, \\
        \hfill \quad \quad \quad \quad \quad \quad \quad \text{if}\ W_{low} \leq I(m,n) < W_{high}\\
      1, \hfill \quad \quad \quad \quad \quad \quad \quad \quad \quad \quad \quad \text{otherwise}
    \end{cases}
  \end{equation}
  where $G(m,n)$ is the final processed image from the proposed framework, $W_{low}$ and $W_{high}$ are tunable parameters (intended for tuning the contrast of the image) such that
  \begin{equation}
      W_{low} < W_{high} < 1 \nonumber
  \end{equation}
 The complete framework of the proposed ultrasound image enhancement and reconstruction process is depicted in Fig. \ref{Block}.
 
\section{Results}
\begin{figure*}[ht!]
\centering
\includegraphics[width = 4.1 in, height = 2.8 in]{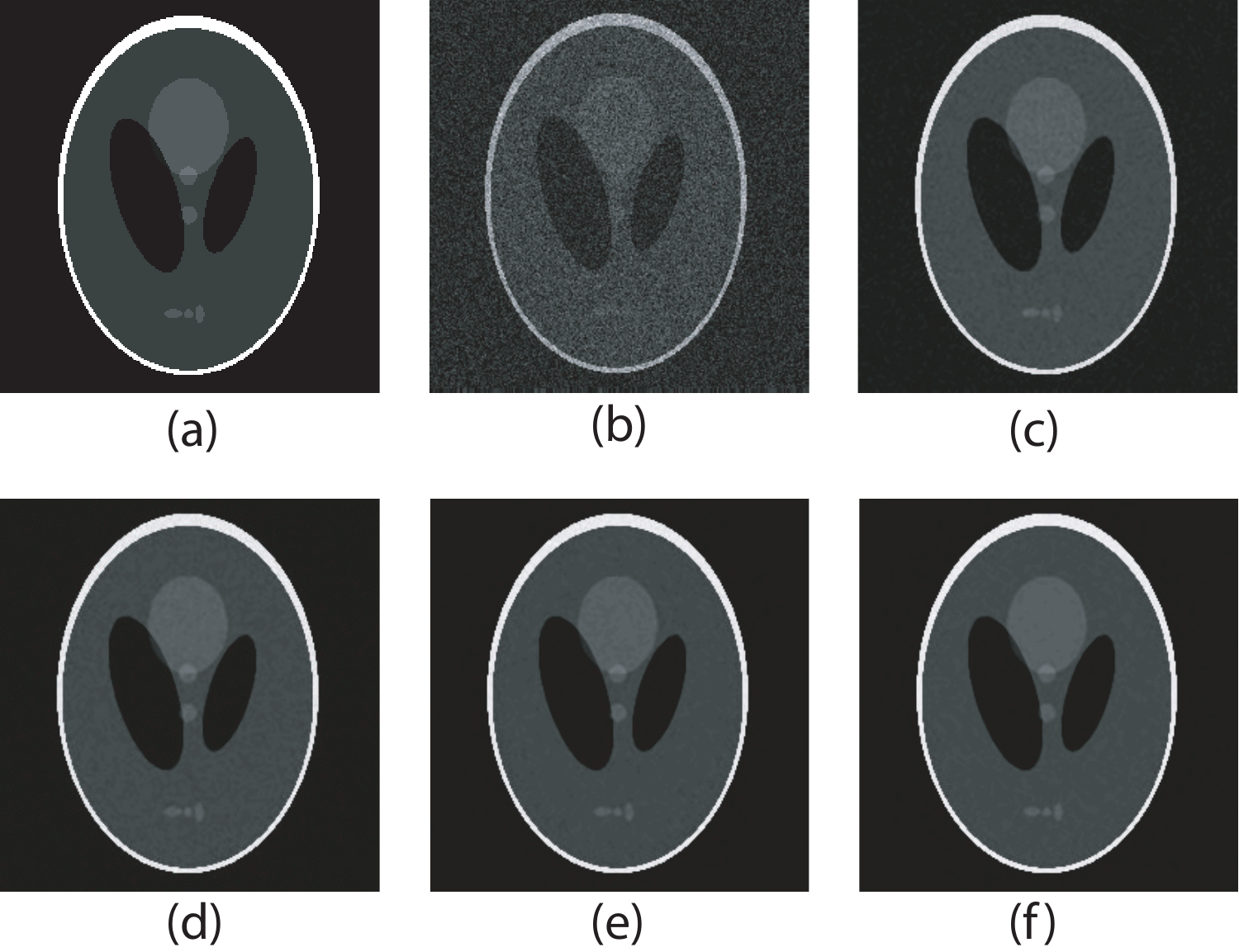}
 \caption{Effect of the number of frames in despeckling the modified Shepp-Logan phantom image  using the proposed algorithm: (a) clean phantom, (b) noisy phantom ($\sigma = 0.4$), (c)-(f) despeckled using 5,10, 15 and 20 frames, respectively.}
 \label{frames}
\end{figure*}

\begin{figure*}[ht!]
\centering
\includegraphics[width = 4.8 in, height = 4.3 in]{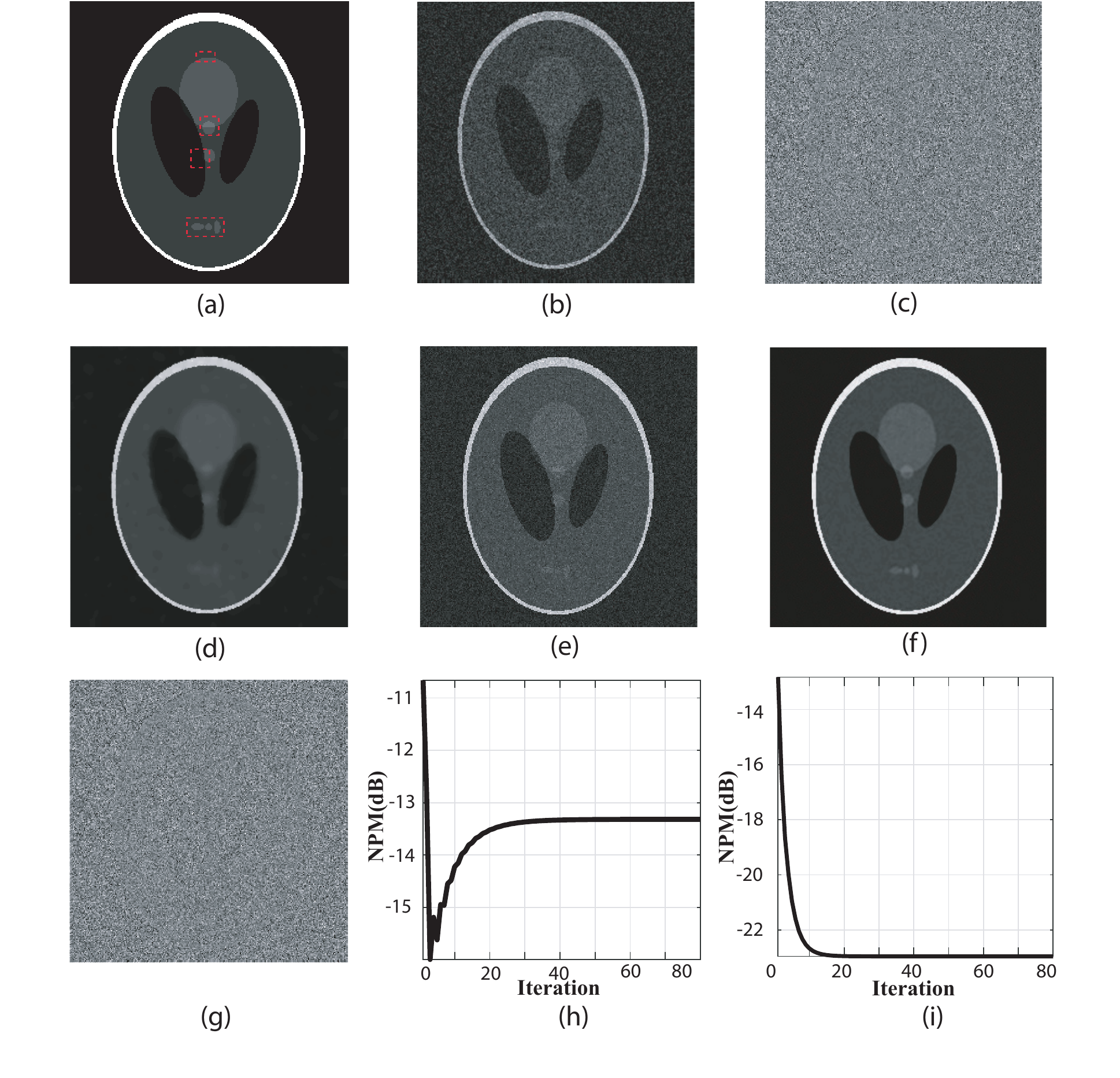}
 \caption{Despeckling of Shepp-Logan phantom image corrupted by synthetic speckle noise. (a) clean phantom with red marked regions considered for EPI calculation, (b) noisy phantom ($\sigma=0.4$), (c) true speckle noise in the $5$th frame; despeckled image using (d) SRAD, (e) OBNLM, (f) proposed MSNE; (g) extracted noise from the $5$th frame using MSNE; (h) NPM measure between the true and estimated noise using MSNE without constraint; (i) NPM measure between the true and estimated noise using MSNE with constraint.}
 \label{phantom}
\end{figure*}
In this section, the efficacy of our proposed framework for high-resolution B-mode image generation is evaluated on both the simulation and \textit{in-vivo} data. The contents of  the paper cover: deconvolution, despecking via MSNE, and post-processing for a complete B-mode image generation. However, as the main contribution is the despeckling algorithm, the simulation study is designed to show the effectiveness of our proposed method for speckle noise estimation and despeckling only. The convergence of the algorithm in the presence of additive noise is also shown to justify the use of the proposed constraint. On the other hand, the \textit{in-vivo} images suffer from low resolution and speckle noise arising from the physical phenomena related with the US image acquisition system. Therefore, according to the signal generation models, the \textit{in-vivo} study includes the $2$-D deconvolution to enhance the resolution as the first phase for all the methods involved for comparing despeckling performance. Two types of deconvolution methods, namely-- \textit{b}MCFLMS and cepstrum \cite{taxt2001two} are investigated. Finaly, post processing stage is included  to make a complete investigation of the high-resolution B-mode image generation pipeline in a single research paper. 

The quality of the despeckled image is compared with those of SRAD (speckle reducing anisotropic diffusion filter) \cite{yu2002speckle} and OBNLM (optimized Bayesian non-local means-based filtering) \cite{coupe2009nonlocal} methods. The performance matrices, used in this case, are SNR (signal-to-noise ratio), PSNR (peak signal-to-noise ratio), SSIM (structural similarity index measure) \cite{shahdoosti2017two}, EPI (edge preservation index) \cite{sattar1997image}, NIQE (natural image quality evaluator) \cite{mittal2013making} and BRISQUE (Blind/Referenceless Image Spatial Quality Evaluator) \cite{mittal2012no}. EPI was calculated for several region of interest near the edges (shown as red marked regions in Fig. \ref{phantom}(a)) and the average EPI is reported. Among the described indices SNR, PSNR, SSIM, EPI require reference image and hence, cannot be used for \textit{in-vivo} data. In the absence of a reference image, NIQE, BRISQUE and visual evaluation are the only ways to evaluate the performance of the proposed algorithm. On the other hand, in case of simulation data with the original noiseless image at hand, we attempt to build the intuition behind different aspects, i.e., the number of image frames to be chosen, runtime, efficacy in preserving small details, of the proposed despeckling algorithm. 

\subsection{Simulation Data} 
Simuation data were generated using the `Modified Shepp-Logan' phantom available in MATLAB with size $256\times256$. In the simulation study, we have investigated the despeckling efficacy of the proposed framework and have not considered the PSF effect on the image. Hence, the image was corrupted with speckle noise only as described in \cite{coupe2009nonlocal} following \eqref{simEqn} where $\mathcal{V}(m,n) \sim \mathcal{N}(0,~\sigma^2)$ and $\eta = 0.5$ was used. The level of noise was varied by setting $\sigma = \{0.2; 0.4; 0.8\}$. At a particular noise level, the speckle pattern was varied using different $\nu(m,n)$ patterns with the same distribution for different frames. Among the frames used for despeckling using our proposed method, the last frame was despeckled using SRAD and OBNLM for comparison with the proposed algorithm. Here, the implementation platform used were: CPU: Intel$\textsuperscript{\textregistered}$ Core$\textsuperscript{TM}$ i$7$-$8700$K, RAM: $32$ GB, software:
MATLAB$\textsuperscript{\textregistered}$, The MathWorks, Natick, MA. Comparing \eqref{speckleModel} and \eqref{simEqn} in additive noiseless case, we can write the speckle pattern of the $k$-th frame as
\begin{equation}\label{speckleNoise}
u_k(m,n) = 1 + r^{-0.5}(m,n)\nu_k(m,n)
\end{equation}
Therefore, \eqref{speckleNoise} can be used to calculate the true speckle pattern for the simulation data, and subsequently, it can be used to quantify the performance of the proposed despeckling algorithm. As claimed in Section \ref{sssec:ref1}, increasing the number of image frames in the proposed speckle pattern estimation algorithm has an impact on the accuracy of estimation. In what follows, we attempt to establish a suitable frame number that optimally meets all the assumptions made in Section \ref{sssec:ref1}, consumes less runtime, and produces a visually pleasant despeckled image. The performance index used in this case is NPM (normalized projection misalignment) defined as
\begin{align}
\mbox{NPM}(q) &= 20\mbox{log}_{10}\left(\frac{\|\rho(q)\|}{\|\mathbf{U}\|}\right)~\mbox{dB} \\
    \rho(q) &= \mathbf{U}-\frac{\mathbf{U}^T\mathbf{\hat{U}}(q)}{\mathbf{\hat{U}}^T(q)\mathbf{\hat{U}}(q)}\mathbf{\hat{U}}(q)
\end{align}
A lower value of NPM indicates better estimation of $\mathbf{U}$. From Table \ref{tab:NPM}, it is apparent that increasing the level of noise deteriorates the estimation accuracy of the speckle pattern. However, it can be improved by around $6$~dB for the three different noise levels as mentioned above by increasing the number of image frames from $5$ to $20$ in the proposed algorithm. As described in \cite{xu1995least}, for an accurate estimation using the blind multichannel algorithm, the channels should not have common zeros. As we introduce more image frames in the estimation process, the probability of having common zeros decreases. This in turn improves the speckle estimation accuracy. However, increasing the number of image frames imply more computational complexity leading to higher runtime, and at the same time, it causes the violation of the quasi-stationarity assumption for the true image as described earlier in Section \ref{sssec:ref1}. The despeckled images using different number of frames for noise level $\sigma = 0.4$ are shown in Fig. \ref{frames}. It is apparent from this figure that consideration of frame number greater than $10$ results in visually imperceptible change in the despeckled images. Therefore, in Fig. \ref{phantom}, we have used $10$ image frames to compare our simulation phantom results with other algorithms.

\begin{table}[h!]
    \caption{Simulation results on the estimation accuracy in terms of NPM (dB) of speckle pattern using the proposed  method for different noise levels}
    \centering
    \begin{tabular}{|c|c|c|c|c|}\hline
    \multicolumn{1}{|c|}{\rule{0pt}{14pt}} & \multicolumn{3}{c|}{\rule{0pt}{14pt}~NPM~(dB)} & \multicolumn{1}{c|}{\rule{0pt}{14pt}} \\
    \cline{2-4}
    \rule{0pt}{14pt}Number of frames  & \multicolumn{3}{c|}{\rule{0pt}{14pt} Noise level} & Runtime (sec)\\
    \rule{0pt}{14pt} & \multicolumn{3}{c|}{\rule{0pt}{14pt}$\sigma=$0.2~~$\sigma=$0.4 ~~$\sigma=$0.8} & \\ \hline
     \rule{0pt}{14pt}$5$ & $-31.65$ & $-25.69$ & $-19.96$ & $0.61$\\
     \hline
     \rule{0pt}{14pt}$10$ & $-34.69$ & $-28.72$ & $-22.96$ & $2.01$\\
     \hline
     \rule{0pt}{14pt}$15$ & $-36.41$ & $-30.45$ & $-24.68$ & $3.95$\\
     \hline
     \rule{0pt}{14pt}$20$ & $-37.64$ & $-31.64$ & $-25.83$ & $6.63$\\
     \hline
    \end{tabular}

    \label{tab:NPM}
\end{table}

\begin{table}
 \caption{Performance measures computed for the simulation study with different noise level ($\sigma$) using diffetrent despeckling approaches}
\centering
\resizebox{\columnwidth}{!}{
\begin{tabular}{|c|c|c|c|c|c|}\hline
 \rule{0pt}{14pt}Methods & Noise level & SNR  & PSNR  & SSIM	& EPI\\
 \rule{0pt}{14pt}  &   ($\sigma$) &    (dB)      &    (dB)    &           &           \\ \hline
 \rule{0pt}{14pt}        & $0.2$ &   $20.33$        &  $24.98$    &  $0.9996$   &  $0.89$\\ \cline{2-6}
 \rule{0pt}{14pt}OBNLM   & $0.4$ &   $14.24$        &  $19.35$    &  $0.9985$   &  $0.80$\\ \cline{2-6}
 \rule{0pt}{14pt}        & $0.8$ &   $10.45$        &  $15.94$    &  $0.9969$   & $0.63$\\ \hline \hline
 \rule{0pt}{14pt}	     & $0.2$ &   $12.97$        &  $18.28$    &  $0.9980$   & $0.90$\\ \cline{2-6}
 \rule{0pt}{14pt}SRAD    & $0.4$ &   $9.32$         &  $15.04$    &  $0.9954$   & $ 0.88$\\ \cline{2-6}
 \rule{0pt}{14pt}	  	 & $0.8$ &   $7.89$         &  $13.62$    &  $0.9935$  & $0.86$\\ \hline \hline
 \rule{0pt}{14pt}        & $0.2$ &   $28.88$        &    $33.22$  &  $0.9999$   & $0.93$\\ \cline{2-6}
 \rule{0pt}{14pt}Proposed MADS & $0.4$ &   $24.30$        &   $28.79$   &  $0.9998$ & $0.91$\\ \cline{2-6}
 \rule{0pt}{14pt}        & $0.8$ &   $21.63$        &   $26.23$   &  $0.9997$  &  $0.89$\\ \hline
\end{tabular}
}

    \label{tab:simComp}
\end{table}

\begin{figure*}[ht!]
\centering
\includegraphics[width = 5.2 in, height = 2.9 in]{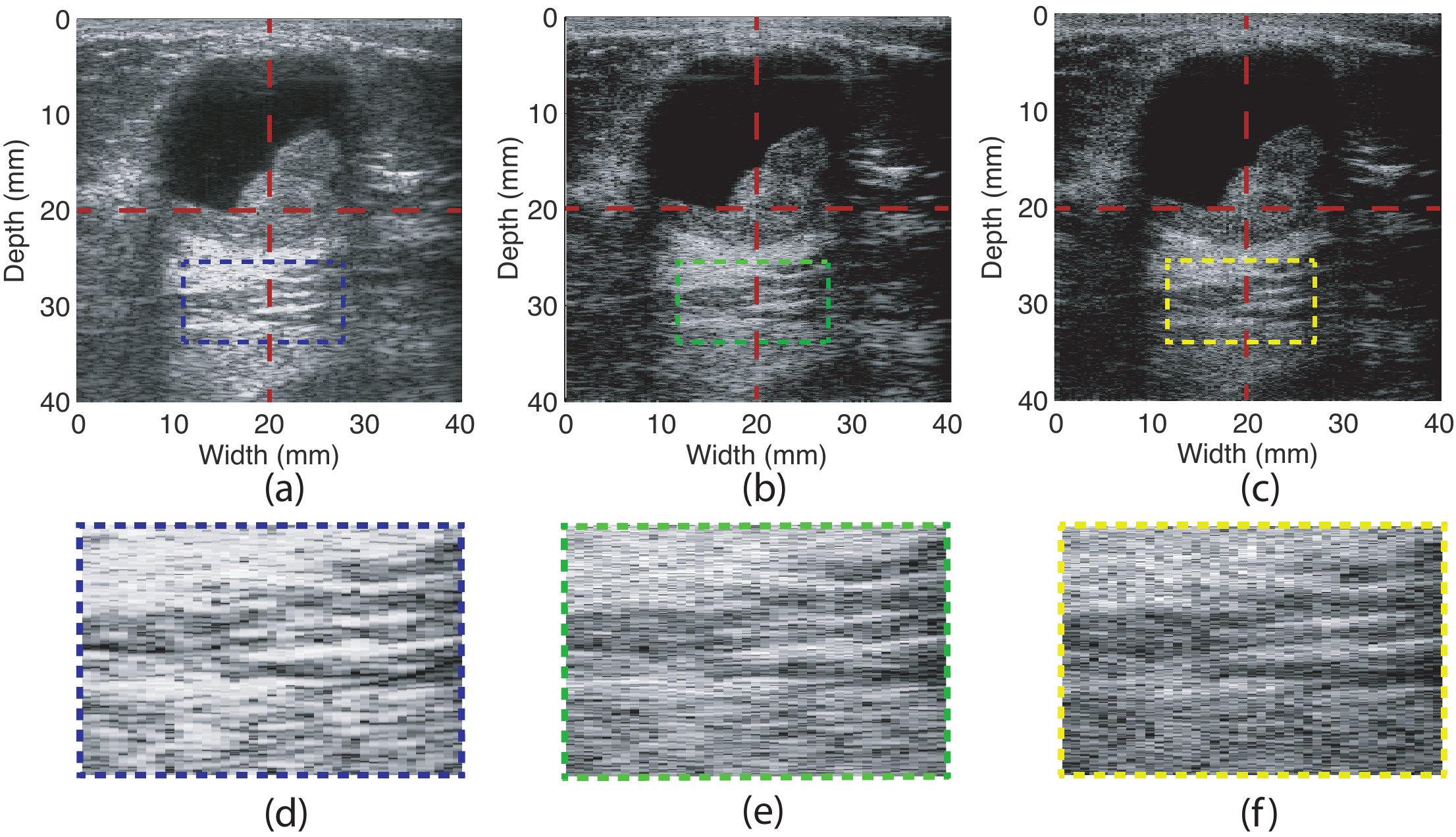}
 \caption{Deconvolution of ultrasound images using adaptive \textit{b}MCFLMS algorithm. (a) Raw RF image, (b) $1$-D deconvolved image, (c) $2$-D deconvolved image, (d)-(f) zoomed-in views of image segments of (a)-(c), respectively.}
 \label{deconv}
\end{figure*}

The performnce of the proposed algorithm in comparison to others is illustrated in Fig. \ref{phantom}. From Fig. \ref{phantom}(d), we can deduce that SRAD distorts the texture in the homogeneous region and blurs the small details in the phantom. Again, as evident from Fig. \ref{phantom}(e), although OBNLM is superior in performance compared to SRAD in preserving texture and edges, it fails to remove speckle noise completely when the noise level is high, e.g., $\sigma = 0.4$. On the other hand, our proposed algorithm (see Fig. \ref{phantom}(f)) shows significant visual improvement in terms of maintaing original texture, edges and small details compared to the SRAD and OBNLM approaches, and the despeckled image is visually close to the clean phantom image. Again, the quantitative metrics as presented in Table \ref{tab:simComp} also demonstrate that our proposed algorithm performs significantly better in terms of quantitative indices SNR, PSNR, and SSIM at different noise levels compared to SRAD and OBNLM. However, SRAD performs nearly as good as MADS in terms of EPI. SRAD was originally designed to enhance the edges without considering the preservation of texture. As a result, SRAD gives nearly similar EPI values and lower values for other performance indices compared to those of MADS. To show the effect of the zero-lag correlation constraint on the convergence profile of the proposed MSNE algorithm, we depict in Fig. \ref{phantom}(h)-(i) the NPM curve with $20$ dB SNR. As can be seen in \ref{phantom}(h), the algorithm misconverges near $-16$ dB, whereas in \ref{phantom}(i), there is no sign of misconvergence and the constrained MSNE algorithm smoothly convergences to around NMP$= -22.96$~dB implying that the zero-lag correlation constraint is effective in preventing misconvergence.

\subsection{\textit{In-Vivo} Data}
Performance of the proposed complete framework for US B-mode image generation comprising of deconvolution, despeckling and post-processing, respectively, is evaluated on the \textit{in-vivo} data, collected using a commercial SonixTOUCH Research (Ultrasonix Medical Corporation, Richmond BC, Canada) scanner integrated with a linear array transducer, L14-5/38, operating at $10$ MHz with sampling frequency of 40 MHz. These data were collected from the patients who appeared for medical examination at the Medical Centre of Bangladesh University of Engineering and Technology (BUET), Dhaka, Bangladesh. This study was approved by the Institutional Review Board (IRB), and prior patient consent was taken. 
 
The performance evaluation of the deconvolution step is done subjectively as elaborate performance evaluation of the $1$-D \textit{b}MCFLMS algorithm is already done in our published work \cite{dey2018ultrasonic}. However, the performance of the complete framework is evaluated using two approaches. First, we keep the deconvolution step fixed for all the despeckling algorithms to be compared with the proposed MADS algorithm and evaluate their comparative performance both from visual and quantitative perspectives. Second, we compare subjectively the image generated using our complete framework with the data acquiring machine B-mode image. All the \textit{in-vivo} images shown here were log compressed and dynamic range was set to $35$ dB as described in \cite{dey2018ultrasonic} for display purpose.

\begin{figure*}[ht!]
\centering
\includegraphics[width = 5.2 in, height = 3.8 in]{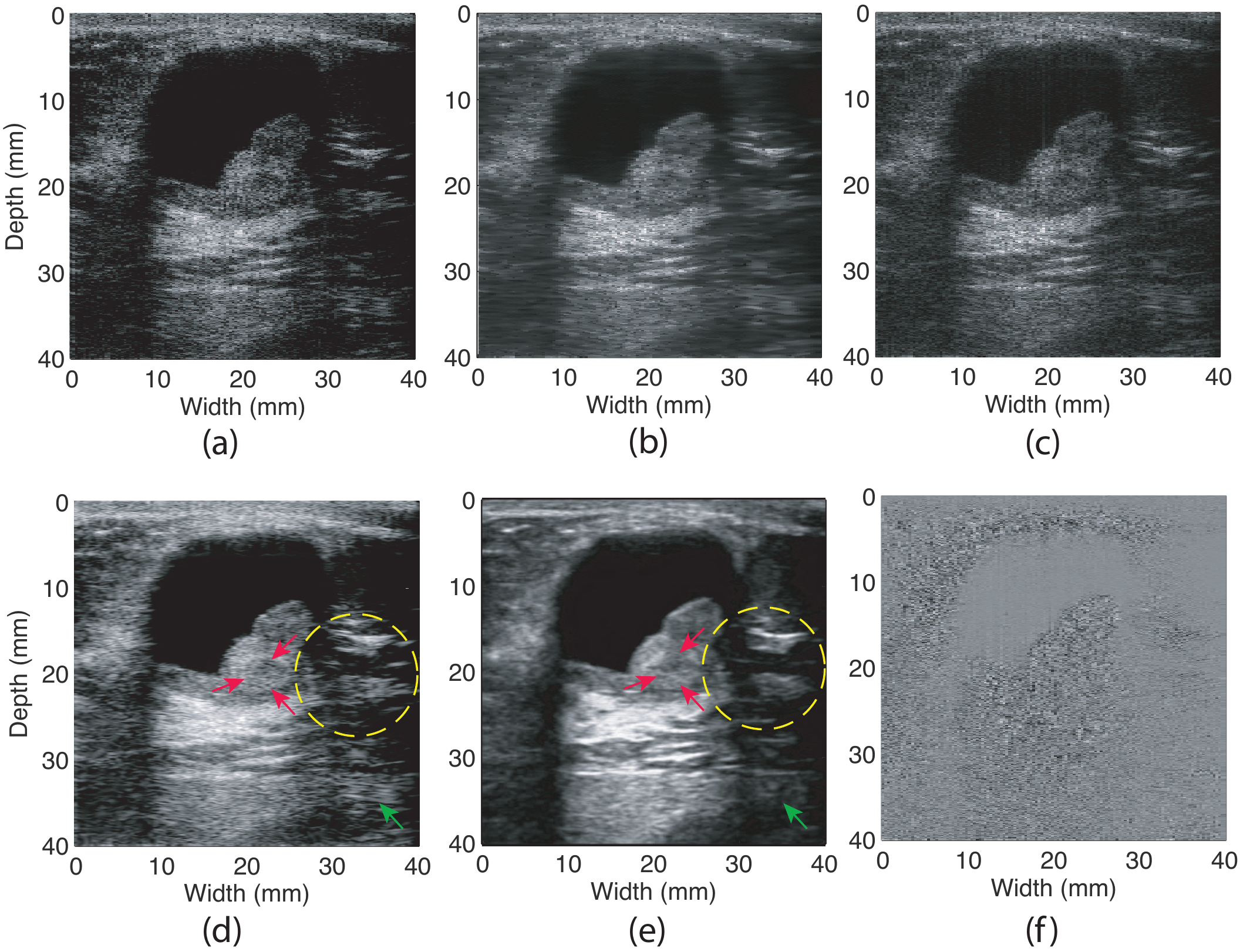}
 \caption{Despeckling of breast ultrasound RF data-1 image. (a) Deconvolved image, images obtaind using (b) SRAD, (c) OBNLM, (d) proposed algorithm, (e) machine B-mode image, (f) estimated speckle pattern of the $5$-th frame.}
 \label{data_1}
\end{figure*}

\begin{figure*}[ht!]
\centering
\includegraphics[width = 5.2 in, height = 3.8 in]{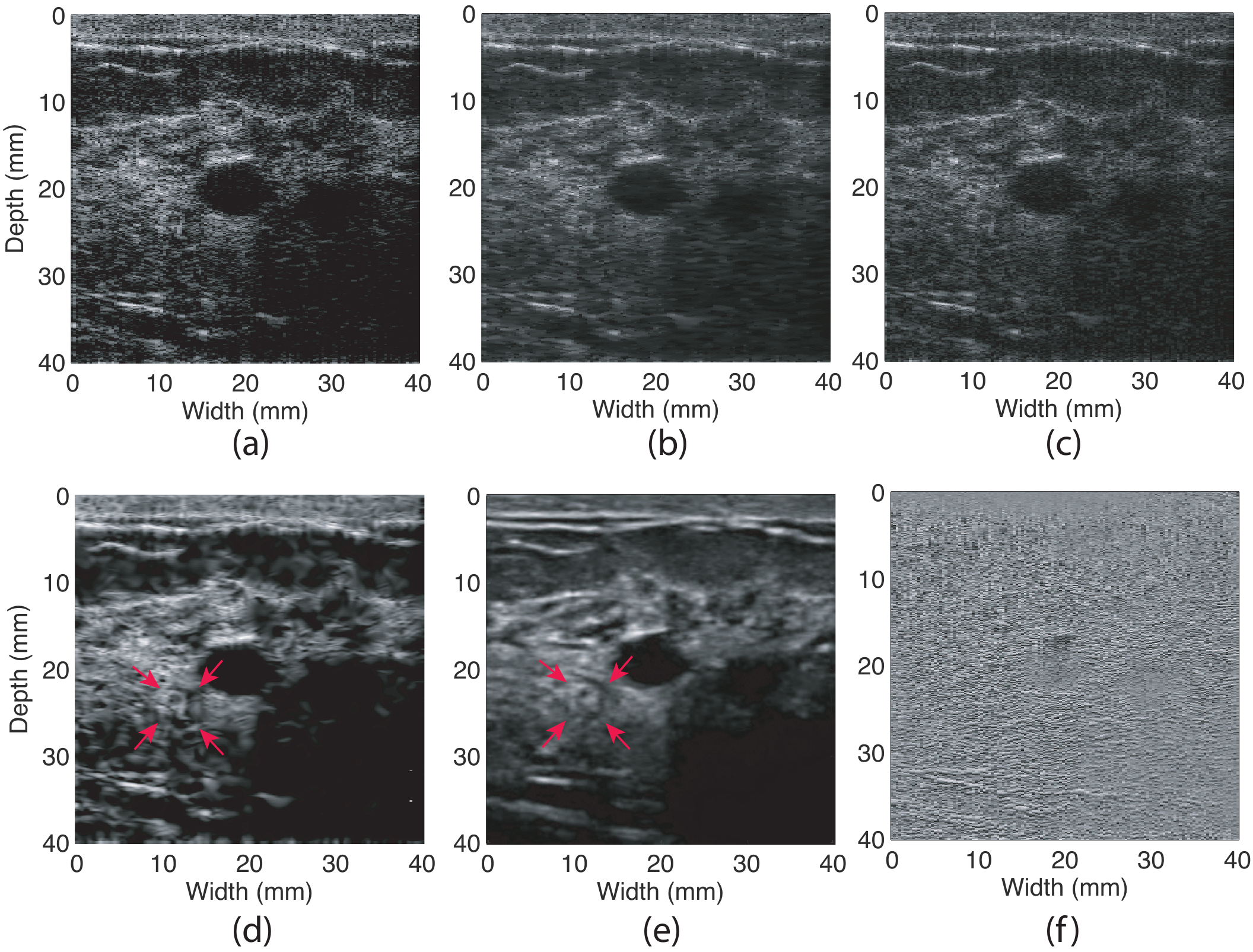}
  \caption{Despeckling of breast ultrasound RF data-2 image. (a) Deconvolved image, images obtaind using (b) SRAD, (c) OBNLM, (d) proposed algorithm, (e) machine B-mode image, (f) estimated speckle pattern of the $5$-th frame.}
 \label{data_2}
\end{figure*}

\begin{table}
 \caption{Axial and lateral correlation energy for raw RF, $1$-D and $2$-D deconvolved data using the b-MCFLMS algorithm.}
\centering
\begin{small}
\begin{tabular}{|c|c|c|}\hline
 \rule{0pt}{10pt} Data & \multicolumn{2}{c|}{\rule{0pt}{10pt}Correlation energy}\\\cline{2-3}
 \rule{0pt}{10pt}      &  Axial   & Lateral \\ \hline
 \rule{0pt}{14pt} RF   & $0.0379$   & $0.0501$  \\ \hline
 \rule{0pt}{14pt} $1$-D deconvolved & $0.0287$ & $0.0497$ \\ \hline
 \rule{0pt}{14pt} $2$-D deconvolved &  $0.0279$ & $0.0329$ \\ \hline
\end{tabular}
\end{small}
    \label{tab:correlation}
\end{table}

\subsubsection{2-D Deconvolution Performance Evaluation on \textit{In-Vivo} Data}
The successive stages of deconvolution offer images with improved and finer texture as shown in Figs. \ref{deconv}(a)-(c) and the zoomed-in views of their marked portion in Figs. \ref{deconv}(d)-(f), respectively. The speckle pattern in Fig. \ref{deconv}(d) is blurry and highly auto-correlated in the spatial domain as convolution of point speckle with the US PSF results in the spreading of the point spatially and thereby, reduces the resolution of the image. This large spatial coverage of speckle leads to considerable correlation between speckle noise not only in the same frame but also in the consecutive frames leading to a greater number of common zeros. To justify our claim, we select an axial and a lateral line from two consecutive frames along the marked lines in Figs. \ref{deconv}(a)-(c) and calculate the energy of the normalized correlation among two axial lines as well as lateral lines of two consecutive frames as presented in Table \ref{tab:correlation}. Higher value of correlation energy indicates higher correlation among the two frames along the axial or the lateral direction. From Table \ref{tab:correlation}, it is evident that the raw RF data of a particular frame is more correlated with the next frame in both the axial and the lateral directions compared to those of the $1$-D and the $2$-D deconvolved data. However, the $1$-D deconvolved data has higher lateral correlation compared to that of the $2$-D deconvolved data with nearly the same axial correlation (see Table \ref{tab:correlation}). And in Fig. \ref{deconv}(e), the speckle pattern becomes fiber-like with greater spatial coverage along the lateral direction than the axial direction leading to higher correlation with the next frame in the lateral direction. Finally, in Fig. \ref{deconv}(f), the lateral correlation is sufficiently removed, and the speckle pattern becomes randomly distributed and uncorrelated. As the speckles in the final $2$-D deconvolved image occupies lesser space, this has additional advantage of reducing common zeros between the speckle patterns in consecutive frames along with increasing resolution of the image. Hence, the speckle pattern estimation can be done efficiently with $5-10$ number of frames without violating the quasi-stationarity assumption of the true image.

\subsubsection{Despeckling Performance on \textit{in-vivo} Data}
\begin{table}[h!]
    \caption{NIQE measure for images despeckled with different algorithms}
    \centering
    \resizebox{\columnwidth}{!}{
    \begin{small}
    \begin{tabular}{|c|c|c|c|c|}\hline 
     \rule{0pt}{14pt}Image & Index & OBNLM & SRAD & Proposed MADS\\
    \hline
    \rule{0pt}{14pt}RF data-1 & NIQE & $7.38$ & $7.69$ & $5.16$\\ \cline{2-5}
    \rule{0pt}{14pt}           & BRISQUE & $34.61$ & $36.44$ & $15.93$\\ \hline
    \rule{0pt}{14pt}RF data-2 & NIQE & $7.61$ & $8.49$ & $5.32$\\ \cline{2-5}
    \rule{0pt}{14pt}           & BRISQUE & $32.27$ & $51.88$ & $19.03$\\ \hline
    \end{tabular}
    \end{small}
    }
    \label{tab:in_vivo}
\end{table}
We have used two \textit{in-vivo} breast ultrasound images, RF image-1 and RF image-2, to validate the performance of our proposed algorithm. Similar type of post-processing with $\gamma = 0.97$ in \eqref{gamma} and $W_{low} = 1e-2, W_{high} = 0.98$ in \eqref{gray} were set for all the images despeckled with different algorithms for illustration purpose. The performance index used here is NIQE that relies on the deviation from the statistical regularities of distortionless images to rate an image as defined in \cite{shahdoosti2018maximum}, \cite{mittal2013making}. The lower the value of the NIQE metric, the better the quality of the despeckled image. However, the NIQE index was measured on the despeckled image without post-processing. As shown in Table \ref{tab:in_vivo}, our proposed algorithm gives the lowest NIQE and BRISQUE score and hence, the best quality image compared to that of SRAD and OBNLM. Figs. \ref{data_1} and \ref{data_2} are provided for subjective evaluation of the proposed algorithm. As shown in Figs. \ref{data_1}(b) and \ref{data_2}(b), SRAD algorithm succeeds in preserving edges although it blurs the texture and degrades the contrast of the image. On the other hand, according to Figs. \ref{data_1}(c) and \ref{data_2}(c) OBNLM shows superior performance compared to SRAD in preserving undistorted texture and contrast. However, it fails to remove the speckle noise completely from the image. To visually compare the performance of our proposed framework to that of the commercial US image acquiring machine used in this experiment, Figs. \ref{data_1}(d)-(e) and \ref{data_2}(d)-(e) are portrayed.
\begin{figure}[ht!]
\centering
\includegraphics[width = 3.2 in, height = 1.6 in]{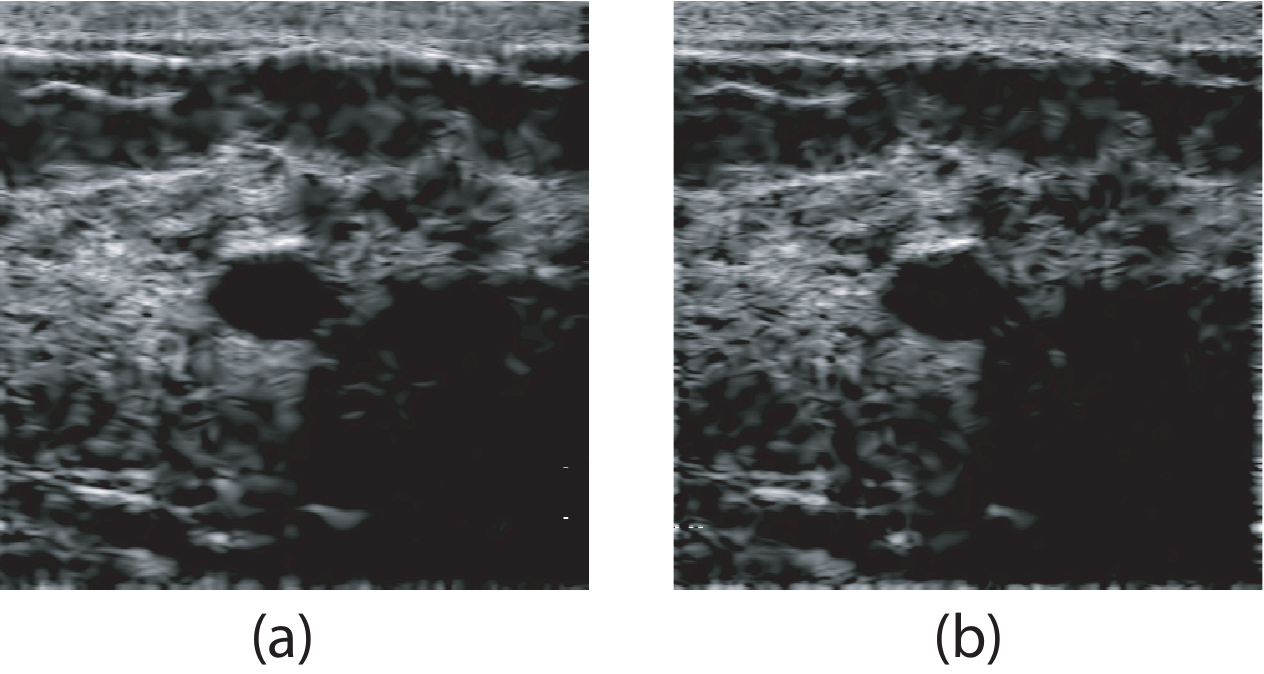}
  \caption{Estimated B-mode image with the deconvolution step as (a) \textit{b}MCFLMS, and (b) cepstrum.}
 \label{cepstrum}
\end{figure}
From these figures, observe that tissue texture is more prominent in the images provided by our proposed framework compared to those in the machine B-mode images. To facilitate the observations, significant structures of the images are marked with arrow and circle in Figs. \ref{data_1}(d)-(e) and \ref{data_2}(d)-(e) which show the machine B-mode images have blurred and distorted the tissue structures. Again, the cyst boundary in Fig. \ref{data_2}(d) is sharper and well-defined comapred to that of Fig. \ref{data_2}(e). The estimated speckle noise patterns of the $5$-th image frame as shown in Figs. \ref{data_1}(f) and \ref{data_2}(f) contain tissue structures that justify the relevance of true image dependent modeling \cite{coupe2009nonlocal} of speckle pattern as shown in \eqref{simEqn}.

In spite of offering an elegant solution to the speckle removal problem, the proposed framework has a flaw in its complete pipeline as the deconvolution step is not realtime implementable requiring $76$ minutes in total for a single image of $128$ A-lines with each line having $1040$ samples. However, as an alternate approach, the deconvolution step can be replaced by a time-efficient cepstrum-based deconvolution as described in \cite{taxt2001two} with a little cost paid in image quality. It is evident from higher NIQE and BRISQUE scores of $5.59$ and $21.23$, respectively using cepstrum deconvolution (see Fig. \ref{cepstrum} (b)) compared to those of $5.32$ and $19.03$, respectively (see Fig. \ref{cepstrum}(a)) using the \textit{b}MCFLMS algorithm, and this brings down the total runtime to $6.3$ seconds. The implementation platform used were: CPU: Intel$\textsuperscript{\textregistered}$ Core$\textsuperscript{TM}$ i5, RAM: 8 GB, software:
MATLAB$\textsuperscript{\textregistered}$, The MathWorks, Natick, MA. A graphical processing unit (GPU) based deconvolution technique to be investigatd in future may bring down the overall B-mode image generation framework into real-time.
\section{Discussion}
In this study, we present a complete framework of signal processing approaches comprising of deconvolution, despeckling, gamma correction, and gray level transformation to produce a high-resolution B-mode image with superior edge and texture from the raw RF image. The parameters for SRAD and OBNLM algorithms were tuned for the lowest NIQE score of the despeckled image. While deriving SRAD and OBNLM algorithms in \cite{yu2002speckle} and \cite{coupe2009nonlocal}, respectively, deconvolution of raw RF image to enhance resolution was not addressed. Hence, introducing deconvolution prior to SRAD and OBNLM may have resulted in their poor performance. In the proposed framework, the performance of the despeckling algorithm (MADS) is dependent on the number of consecutive image frames to be considered in the MSNE algorithm. As mentioned earlier, there is a tradeoff between the number of frames that can be used without violating the quasi-stationarity assumption of the true US image and the speckle noise estimation accuracy. In our experiment, we observed that five consecutive image frames are good enough for a visually pleasant B-mode image generation. Again, the Lagrange multipliers-- $\beta_1$ in the constraint preventing misconvergence of the MSNE algorithm (see \eqref{constraint2}) and $\beta_2$ in the energy constraint of the iterative despeckling algorithm (see \eqref{modifiedCost}) remain effective once set at an optimum level for a particular US imaging set-up. To make the framework independent of the display monitor, gamma correction as a post-processing step has been introduced. Again, to offer the user a tunable contrast adjustment, two parameters $W_{low}$ and $W_{high}$ have been used in the gray level transformation step. 

In addition to providing a guideline for high-resolution B-mode image generation, the paper introduces a method to extract the speckle pattern inherent in a US image. Despite of being a random process, speckle noise is not devoid of information. Since the statistics of the speckle depends on the microstructure of the tissue parenchyma, it can be useful for differentiating between different tissue compositions or types \cite{wagner1983statistics}, \cite{sehgal1993quantitative}.

\section{Conclusion}
This paper has dealt with a complete framework for high-resolution ultrasound image reconstruction from raw RF data. The proposed framework relies on SIMO models for both deconvolution and speckle noise estimation, and MISO model for despeckling. In the first step, to enhance the resolution, a $2$-D deconvolution technique has been introduced as an extension of our previously proposed $1$-D \textit{b}MCFLMS algorithm which is necessary prior to despeckling according to the mathematical model of US imaging. In the next step, a novel multiframe-based adaptive speckle noise estimation (MSNE) algorithm estimates the speckle pattern without any \textit{a priori} information on the statistics of the image or the noise pattern. Using the estimated speckle pattern, an energy constrained iterative algorithm estimates the true US image following a MISO model. As the despeckling procedure is completely based on signal generation model and does not involve any kind of ad-hoc filtering operation as reported in the literature, it has resulted in a high quality tissue texture and edges in the image. Finally, gamma correction and gray level transformation have been done as post-processing to complete the pipeline of high quality B-mode image generation. The results have demonstrated the superiority of our proposed despeckling algorithm compared to SRAD and OBNLM methods. Again, the proposed framework offers B-mode image with superior texture and image details compared to those provided by a commercial ultrasound scanner.

As our proposed framework preserves original image features such as texture, details and edges, it may have a far reaching impact on medical imaging for diagnostic purpose. At the same time, the proposed despeckling algorithm may be efficacious in dealing with the speckle noise problem in other imaging such as synthetic aperture radar (SAR) \cite{wang2019bayesian} and optical coherence tomography (OCT) \cite{paul2019speckle}.

\ifCLASSOPTIONcaptionsoff
  \newpage
\fi
\section*{Acknowledgment}
This work has been supported by the Higher Education Quality Enhancement Program (HEQEP) of University Grants Commission (UGC), Bangladesh (CPSF-96/BUET/W-2/2017).


\ifCLASSOPTIONcaptionsoff
  \newpage
\fi

\bibliographystyle{IEEEtran}

\bibliography{jd}

\newpage

\begin{IEEEbiography}[{\includegraphics[width=1 in,height=1.2in]{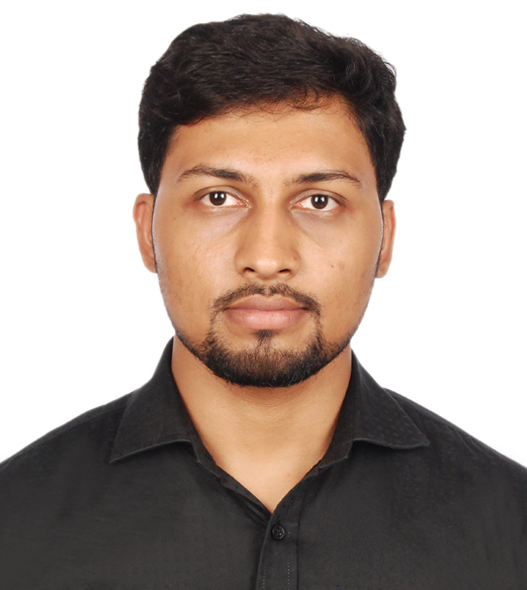}}]{Jayanta Dey}
received the B.Sc. and M. Sc. degrees in electrical and electronic engineering from the Bangladesh University of Engineering and Technology (BUET), Dhaka, Bangladesh, in 2017 and 2019, respectively. He is currently pursuing his PhD in the BME dept. of Johns Hopkins University.

He was a participant at IEEE Signal Processing Cup 2016, Shanghai, China and his team achieved the
1st position at that competition. His current research interests include digital signal processing, adaptive filtering, image processing, neuroscience, machine learning, and medical imaging.
\end{IEEEbiography}


\begin{IEEEbiography}[{\includegraphics[width=1in,height=1.2in]{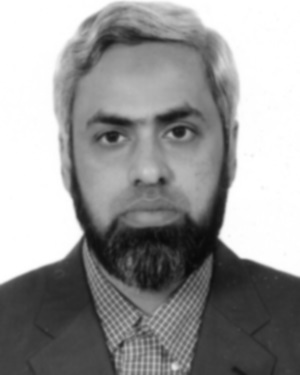}}]{Md. Kamrul Hasan}
received the B.Sc. and M. Sc. degrees in electrical and electronic engineering from the Bangladesh University of Engineering and Technology (BUET), Dhaka, Bangladesh, in 1989 and 1991, respectively, and the M.Eng. and Ph.D. degrees in information and computer sciences from Chiba University, Chiba, Japan, in 1995 and 1997, respectively.

In 1989, he joined BUET as a Lecturer with the Department of Electrical and Electronic Engineering, where he is currently working as a Professor. He was a Postdoctoral Fellow and a Research Associate at Chiba University and Imperial College, London, U.K., respectively. He was a short-term invited Research Fellow at the University of Tokyo, Tokyo, Japan, and a Professor of International Scholars of Kyung Hee University, Seoul, Korea. He has authored or coauthored more than 100 scientific publications. His current research interests include digital signal processing, adaptive filtering, image processing, quantitative ultrasound, elastography, machine learning, and medical imaging. Dr. Hasan is currently an Associate Editor for the IEEE Access.
\end{IEEEbiography}

\end{document}